\documentclass[twocolumn]{aastex63}

\shorttitle{Disconnecting the inner and outer solar system}
\shortauthors{Izidoro et al.}
\graphicspath{{./}{figures/}}

\usepackage{amsmath} 
\usepackage[T1]{fontenc}
\usepackage[utf8]{inputenc} 

\begin{document}

\title{The effect of a strong pressure bump in the Sun's natal disk: Terrestrial planet formation via planetesimal accretion rather than pebble accretion}

\correspondingauthor{André Izidoro}
\email{izidoro.costa@gmail.com}

\author{André Izidoro}
\affiliation{Department of Earth, Environmental and Planetary Sciences, Rice University \\ 6100 Main Street, MS 126, Houston, TX 77005, USA }

\author{Bertram Bitsch}
\affiliation{Max-Planck-Institut für Astronomie 
Königstuhl 17, 69117 Heidelberg, Germany}

\author{Rajdeep Dasgupta}
\affiliation{Department of Earth, Environmental and Planetary Sciences, Rice University \\ 6100 Main Street, MS 126, Houston, TX 77005, USA }





\begin{abstract}


Mass-independent isotopic anomalies of carbonaceous  and non-carbonaceous meteorites show a clear dichotomy suggesting an efficient separation of the inner and outer solar system. Observations show that ring-like structures in the  distribution of mm-sized pebbles in  protoplanetary disks are  common. These structures are often associated with drifting pebbles being trapped by local pressure maxima in the gas disk. Similar structures may also have existed  in the sun's natal disk, which could naturally explain the meteorite/planetary isotopic dichotomy. Here, we test the effects of a strong pressure bump in the outer disk (e.g. $\sim$5~au) on the formation of the inner solar system.  We model dust coagulation and evolution, planetesimal formation, as well as embryo's growth via planetesimal and pebble accretion. Our results show that terrestrial embryos formed via planetesimal accretion rather than pebble accretion. In our model, the radial drift of pebbles foster planetesimal formation. However, once a pressure bump forms, pebbles in the inner disk are lost via drift before they can be efficiently accreted by embryos growing at $\gtrapprox$1~au. Embryos inside $\sim$0.5-1.0au grow relatively faster and can accrete pebbles more efficiently. However, these same embryos grow to larger masses so they should migrate inwards substantially, which is inconsistent with the current solar system. Therefore, terrestrial planets most likely accreted from giant impacts of Moon to roughly Mars-mass planetary embryos formed around $\gtrapprox$1.0~au. Finally, our simulations produce a steep radial mass distribution of planetesimals in the terrestrial region which is qualitatively aligned with formation models suggesting that the asteroid belt was born low-mass.



\end{abstract}

\keywords{terrestrial-planets --- Earth --- Mars --- asteroids --- meteorites}


\section{Introduction} \label{sec:intro}


Protoplanetary disks are highly dynamic environments where gas strongly influences the motion of the solids. Sufficiently small dust grains are dynamically well coupled to the gas and largely follow the gas motion~\cite[]{weidenschilling77,braueretal08}. Due to the vertical component of the stellar gravitational acceleration, dusts tend to settle towards the gas disk midplane~\citep[e.g.][]{nakagawaetal81,nakagawaetal86}. In weakly turbulent disks, small dust grains eventually grow to millimeter and centimeter size particles~\citep[e.g.][and references therein]{blumwurm00,birnstieletal16,blum2018,teiseretal21}. These particles are usually refered as pebbles~\citep[e.g.][]{johansenlambrechts17}.

In an idealized gaseous protoplanetary disk where the  gas pressure decreases monotonically as a function of orbital distance, the gas  rotates around the central star at sub-keplerian speed everywhere in the disk~\cite[e.g.][]{adachietal76}. In such a disk, pebbles traveling around the star at Keplerian speed feel a head-wind and may drift all the way from the outermost parts of the disk until they may get accreted into the star or recycled in the disk \cite[e.g.][]{armitage10}. Pebble-drift timescales are much shorter than the  gas disk lifetimes~\cite[e.g.][]{weidenschilling77,braueretal08,willianscieza11}.





While drifting, pebbles may be spontaneously concentrated in specific regions of the disk and promote the formation of planetesimals via gravitational instabilities in over-dense pebble clumps \citep[e.g.][]{youdinshu02,cuzziweidenschilling06,Johansenetal09,klahretal18}. Planetesimals grow to protoplanetary embryos that further grow to planets via accretion of drifting pebbles and planetesimal-planetesimal accretion \citep[e.g.][]{johansenlambrechts17}. 


When growing planets reach masses of $\sim$10-20 Earth masses, they gravitationally interact with the surrounding gas and change the disk structure creating  a high pressure region in the gas disk outside its orbit~\citep[these structures are typically refereed as pressure bumps; ][]{paardekoopermellema06,morbidellinesvorny12,lambrechtsetal14,bitschetal18b}. In the pressure bump, the gas of the disk speeds up, potentially reaching super-Keplerian velocities. Consequently, inward drifting  pebbles reaching this region do not necessarily continue to drift inwards because gas-drag effects tend to vanish at the pressure maxima~\citep[e.g.][]{weidenschilling80,bitschetal18b}. Sufficiently large pebbles are mostly trapped at the pressure bump, which drastically reduces or even completely shuts down pebble accretion on the growing planet inside the pressure bump.




For the Solar System, it has been proposed that Jupiter's early formation induced a pressure bump in the sun's natal disk that regulated the pebble flux to the terrestrial region~\citep{lambrechtsetal19}. In this scenario, Jupiter was responsible for separating the Solar System inner and outer parts. Mass-independent, isotopic anomalies among carbonaceous (hereafter referred as CC) and non-carbonaceous (hereafter referred as NC) meteorites support this scenario. A distinct difference in stable isotopic composition between these two classes of meteorites has been documented for a wide range of non-volatile elements such as siderophile Mo, W, Ni, Ru and lithophile Cr, Ti \citep{warrenetal11,Kruijeretal17,kruijeretal20}. Stable isotopic compositional dichotomy between NC and CC meteorites has recently been extended to volatile element nitrogen~\citep[N; ][]{grewaletal21}. Combined Cr-Ti-O isotope data also suggests no transport of chondrules -- a particular type of pebbles~\citep[see][for a review on chondrules formation models]{connollyetal16} -- between the NC and CC reservoirs~\citep{schneideretal20}.

In a nominal gas disk, Jupiter's core growth via pebble accretion may not have been fast enough to prevent significant mixing of the inner and outer solar system pebble reservoirs~\citep{brassermojzsis20}. It may well take more than 0.5-1~Myr for Jupiter to grow to masses where it can significantly block the pebble flux from the outer disk. However, this late formation may not be a  problem if  all pebbles drifting into the inner disk before Jupiter's core growth (first 0.5-1~Myr) had   a NC-like composition~\citep[e.g.][]{spitzeretal20}. If this is not the case,  another mechanism may be required to early disconnect the inner and outer solar system reservoirs -- before Jupiter's core formation -- and avoid the mixing of the NC and CC pebble reservoirs.

One such possibility  is the existence of an intrinsic pressure bump in the disk~\citep{brassermojzsis20}. Pressure maxima in gaseous disks are indeed observed in hydrodynamical simulations, for instance, at the disk water snowline\footnote{Distance from the star where water condensates as ice.}~\citep{{bitschetal15,savvidouetal20}}. This bump is caused by different dust grain sizes inside and outside the snowline, and thus an opacity transition at this location~\citep{mulleretal21}. Magnetohydrodynamic hydrodynamical simulations also show pressure bumps due to zonal flows in the disk~\citep[e.g.][]{johansenetal09b,simonarmitage14,flocketal15}. A pressure bump in the disk may in fact accelerate planet formation and produce a giant planet core in a timescale as short as $\sim$10~kyr \citep[e.g.][]{morbidelli20,guilleraetal20} at 5~au, which makes this scenario even more appealing.




From an observational perspective, this idea is also attractive because observations of  protoplanetary disks at mm-wavelengths have shown features consistent with the emission from small and large dust particles (or pebbles) in the form of (multiple) ring-like structures~\citep{alma15,andrewsetal18,isellaetal18}. These structures seem to be present in most observed disks -- if not all -- with some disks showing multiple rings. This suggests that dust and pebbles have been concentrated in specific locations~\citep{dullemondetal18} of the disk potentially associated to pressure maxima in the gas disk. The very mechanism promoting the formation of these annular structures is heavily debated. It has been proposed that these rings are by-products of planet-disk gravitational interaction~\citep[e.g.][]{papaloizoulin84,nelsonetal00,zhuetal14}, but other hypotheses to explain their origins also exist. These include dust pile-up: 1) at snowlines of different molecular species~\citep[e.g.][]{banzattietal15,zhangetal15,pinillaetal17}; 2) due to coupling between magnetic fields and disk material~\citep[e.g.][]{baietal14,simonarmitage14}; 3) at deadzones~\cite[e.g.][]{flocketal15,lyraetal15}; and 4) due to instabilities arising from dust-gas interaction~\cite[e.g.][]{takahashietal14,dullemondpenzlin18,tominagaetal20}. Although ALMA observations have targeted mostly class-II proto-stars ($\sim$ 1~Myr old) even much younger stars exhibit such prominent features in dust emission~\citep{sheehaneisner18,Seguracoxetal20}. Therefore, a key question is whether the Solar System gas disk had a ring due to an early formed pressure bump? If it did, what would be the consequences of such an intrinsic pressure bump on the formation of planetesimals and inner planetary embryos? The goal of this paper is to study the growth mode of terrestrial planetary embryos in disks with pressure bumps in the outer disk. In our simulations, we assume that the envisioned pressure bump forms early, in $\leq$1~Myr after disk's formation.

In our simulations we assume that the location of the pressure bump is associated with the disk water snowline. Once the bump forms, we assume that all dust at that moment inside the pressure bump has a NC-like composition and that outside it is CC-like. We model the effects of dust coagulation, planetesimal formation, and pebble and planetesimal accretion in a gaseous protoplanetary disk. By modeling the growth of planetary embryos from planetesimals and considering the disconnection of the inner and outer pebble reservoirs, our model represents a significant improvement compared to previous studies~\citep{lambrechtsetal19,brassermojzsis20}. 

Our results show that if the pressure bump forms early in the disk -- potentially triggering the prompt formation of Jupiter's core -- and leads to an efficient separation of the inner and outer solar system pebble reservoirs, then terrestrial planetary embryos were most likely formed via planetesimal accretion rather than pebble accretion. The planetesimal formation efficiency in our model is treated as a free-parameter. If planetesimal formation efficiency is sufficiently high near 0.5~au, terrestrial planetary embryos  grow to large masses and should migrate to the disk inner edge becoming hot planets/super-Earths. All our simulations form planetesimals inside 0.4-0.5~AU, which is probably inconsistent with the current Solar System. We envision two potential solutions for this issue: i) the Solar System natal disk was hotter compared to our nominal disk model such that dust grains sublimated inside $\sim$0.5~au; ii) our assumed conditions for planetesimal formation to occur via gravitational collapse of over-dense clumps -- created via zonal flows, vortices and streaming instability -- are too generous and  in fact it never took place inside $\sim$0.5~au because of local low Stokes number and/or dust-to-gas ratio.

This paper is structured as follows. In section \ref{sec:model} we present the dust coagulation and gas disk models used in our simulations. In section \ref{sec:results} we present the our results. In section \ref{sec:discussion} we discuss the implications of our study and its limitations. Finally, in section \ref{sec:conclusions} we summarize the main findings of our study. 

\section{Model} \label{sec:model}

We model the radial drift, coagulation, fragmentation, and turbulent mixing of dust grains in a 1D gaseous protoplanetary disk. Our code is based on the ``Two-pop-py''  code \citep{birnstieletal12}. The code computes the evolution of two dust-size populations -- namely the largest and smallest dust particles -- rather than the  solving dust evolution for a ``continuum'' of dust grain sizes. Note that, in reality, as dust grains evolve in the disk -- by growing and fragmenting -- a dust size distribution is created with grain sizes varying between those of the largest and smallest possible dust particles. Our approximation is fairly decent for our purpose because as dust grains grow, most of the mass is carried  by the largest dust grains \citep[e.g.][]{birnstieletal10}. A similar approach has been invoked in several studies of dust coagulation ~\citep[e.g.][]{drazkowskaetal16,lenzetal19,uedaetal19,voelkeletal20} and pebble accretion~\cite[e.g.][]{lambrechtsetal14,izidoroetal19,bitschetal18c,bitschetal19}. The main advantage of using this approach is its low computational cost compared to solving the evolution of a ``continuum'' of dust grain sizes~\citep{birnstieletal12,birnstieletal15}. We have compared the results of our code with those in  \cite{pinillaetal12}, who follow the evolution of  multiple dust size populations, and found good agreement overall. 


\subsection{Gas disk model}

The gas disk surface density profile in  our simulations is based on the Minimum mass solar nebula model \citep{weidenschilling77,hayashi81}. The central star is a solar-mass star. For simplicity, the gas disk surface density is kept constant over time in most of our simulations~\cite[see also][for a similar approach]{lenzetal19}. This is justified because we are mostly interested on early evolution of the disk, namely, during its first 1~Myr. For completeness, in order to understand the impact of the gas surface density on our results we also present simulations considering an end-member scenario where the gas disk is initially very low mass. The gas disk extends from about 0.35~au to 350~au in all our simulations, which is consistent with the typical sizes of observed disks~\citep{andrewsetal18,andrewswilliams07}.

We have performed simulations including and neglecting the effects of one pressure bump in the disk. To mimic the presence of the bump we modify our original gas disk profile by invoking a simple rescaling function \citep{pinillaetal12,dullemondetal18,morbidelli20}.


In disks with no pressure bumps the vertically integrated gas density is represented by the so called ``Minimum mass solar nebula model''~\citep[MMSN;][]{weidenschilling77,hayashi81}
\begin{equation}
\Sigma_{\rm mmsn}(r) = 1700 \left(\frac{r}{1~{\rm au}}\right)^{-1.5} {\rm \frac{g}{cm^2}},
\label{eq:mmsn}
\end{equation}
where $r$ represents the heliocentric distance.
In disks with a pressure bump, the power-law gas surface density profile is modified to mimic the local perturbation of the gas as 
\begin{equation}
\Sigma_{\rm pb}(r) = \Sigma_{\rm mmsn}(r)  \exp\left[-A\exp\left(-\frac{(r-r_{\rm pb})^2}{w(r_{\rm pb})^2}\right)\right],
\label{eq:1pb}
\end{equation}
\citep{pinillaetal12,dullemondetal18,morbidelli20} where $r_{\rm pb}$, $A$, and $w$ represent the pressure bump radial location, amplitude, and width, respectively. We will perform simulations with and without pressure bumps, and in the latter cases assuming that it appears at different times during the disk's life. 


In all our simulations, the  pressure bump is considered to be a static structure, which is certainly very simplistic. However, as we are mostly interested in understanding how a strong pressure bump influences planet formation in the inner disk, we consider this approximation acceptable. We envision that our long-lived and azimuthally extended pressure bump is either caused by planets or disk physics and chemistry. Following \cite{pinillaetal12} and \cite{dullemondetal18} we set
\begin{equation}
w(r_{\rm pb})=fH_{\rm gas}(r_{\rm pb}),
\end{equation} where $f$ is a dimensionless parameter of order of unit, and $h_{\rm gas}(r)= H_{\rm gas}(r)/r = 0.037\left(\frac{r}{1au}\right)^{2/7}$ is the gas disk aspect ratio~\cite[e.g.][]{chianggoldreich97,chambers09}. $H_{gas}$ is the disk scale height.


For simplicity, the radial gas disk  temperature is also kept constant in  our nominal simulations\footnote{We have also performed a few simulations where the disk dissipates and its temperature drops following an exponential decay with e-fold timescale of 1~Myr and our results did not change qualitatively. }, and it is given as
\begin{equation}
T(r) =  h_{\rm gas}^2 \frac{G M_{\odot}}{r}\frac{\mu}{\mathcal{R}},
\label{eq:3}
\end{equation}
where  $\mathcal{R}$ is the ideal gas constant, $\mu$ is  the gas mean molecular weight set equal to 2.3~${\rm gmol^{-1}}$, $G$ is the gravitational constant, $M_{\odot}$ is the solar mass. In our nominal disk, the disk snowline is at about 5~au. Having the disk snowline at 5~au means that our disk is  hotter than classical passive disks~\citep{chianggoldreich97,chambers09}. Our disk temperature is more  consistent with models of actively acreting young disks~\cite[e.g.][]{minetal11,bitschetal15}.


The gas pressure at the disk's midplane is calculated in the locally isothermal limit as
\begin{equation}
P(r) = c_{\rm s}\frac{\Sigma_{\rm x}}{H_{\rm gas}\sqrt{2\pi}},
\end{equation}
where the sound speed is $c_{\rm s}(r) = H_{\rm  gas}/\Omega_{\rm k}$, and $\Omega_{\rm k}(r) = \sqrt{GM_{\odot}/r^3}$ represents the Keplerian orbital frequency. $\Sigma_{\rm x}$ takes the form of one of our different gas disk profiles (e.g. Eq. \ref{eq:mmsn} or \ref{eq:1pb}).

Long-lived pressure bumps can not be narrower than about the local disk pressure scale height \citep{yangmenou10,onoetal16}. A pressure bump narrower than the disk scale height would probably imply that the disk has not reached hydro-static equilibrium yet. In this case, Rossby wave instability would be potentially triggered and the axial symmetry of the pressure bump would be lost \citep{dullemondetal18}. In light of these arguments, in our simulations we restrict  $w$ to values where $1\leq f \leq3$ \citep{pinillaetal12}. The amplitude of the pressure bump in our simulations is considered to be a free parameter, but we restrict the values of $A$ to be roughly consistent with the density oscillations due to zonal flows driven by magneto-hydrodynamical instabilities \citep{uribeetal11} or gaps opened by giant planets~\cite[e.g.][]{cridaetal06}.

The radial gas velocity is calculated via the following equation, which corresponds to the viscous evolution of the disk \citep{lyndenbellpringle74}
\begin{equation}
v_{\rm r,gas} = -\frac{3}{\Sigma_{\rm x}\sqrt{r}}\frac{\partial}{\partial r}(\Sigma_{\rm x}\nu\sqrt{r}).
\label{eq:vrgas}
\end{equation}
In Eq. \ref{eq:vrgas}, the gas disk viscosity is calculated within the $\alpha$-viscosity paradigm~\citep{shakurasunyaev73} where
\begin{equation}
\nu=\alpha c_s h_{gas}.
\end{equation}
$\alpha$ represents the Shakura-Sunayev  viscosity parameter. In order to calculate the radial velocity of the gas, we use finite difference approximation between adjacent points of our radial grid to compute the derivative in Eq. \ref{eq:vrgas} (also in  Eq. \ref{eq:eta}).

The azimuthal velocity of the gas is characterized by the pressure support parameter
\begin{equation}
\eta = -\frac{1}{2}\left( \frac{c_{\rm s}}{v_{\rm k}}\right)^2\frac{\partial \ln{P}}{\partial \ln{r}}, 
\label{eq:eta}   
\end{equation}
where $v_{\rm k}$ is the keplerian velocity.

\subsection{Dust surface density evolution}

Our nominal gas disk model assumes from the beginning a solar dust-to-gas ratio everywhere in the disk, namely $Z=1.5\%$ \citep[e.g.][]{asplundetal09}. In our simulations, the initial dust surface density is represented by
\begin{equation}
\Sigma_{\rm dust} = 0.015 \Sigma_{mmsn}.
\end{equation}
At the beginning of our nominal simulations, the total mass in dust in the inner ($<$5~au) and outer disk ($>$5au) is about 18 and 180$M_{\rm Earth}$, respectively. Note that the solar system metallicity calculated by \cite{asplundetal09} includes several elements heavier than He and many species that do not condensate inside the disk snowline (e.g. ${\rm H_2O}$, CO, ${\rm N_2}$, etc).  Thus, the very initial dust-to-gas ratio inside the snowline could be a factor of 2-3 smaller than that considered here. We explore the impact of the initial dust-to-gas ratios on our results in a limited number of  simulations (see section \ref{sec:effectgas}). These simulations indicate that invoking a lower initial dust-to-gas ratio inside the snowline does not affect the main, broad conclusions of our work.

To follow the radial evolution of the dust we solve the one-dimmensional continuity   equation for the column dust density \citep{birnstieletal10}
\begin{equation}
\begin{split}
\frac{\partial{\Sigma_{\rm dust}}}{\partial{t}} & +  \frac{1}{r}\frac{\partial}{\partial{r}}\left\{ r \left[\overline{v}_{\rm r,dust}\Sigma_{\rm dust} -  D_{\rm dust}  \frac{\partial}{\partial r }\left(\frac{\Sigma_{\rm dust}}{\Sigma_{\rm x}}\right)\Sigma_{\rm x} \right]  \right\} =  \\  =  \frac{\partial \Sigma_{\rm pla}}{\partial t},
 \label{eq:continuity}
\end{split}
\end{equation}
where the dust diffusivity
\begin{equation}
D_{\rm dust} =  \frac{\alpha c_{\rm s} H_{gas}}{1+St^2},    
\end{equation}
and $\overline{v}_{\rm r,dust}$, which is the mass weighted radial velocity of dust, is calculated following \cite[][see their Eq. 24]{birnstieletal12}. $\Sigma_{\rm pla}$ represents  the planetesimal surface density. We test the effects of different gas viscosities in our simulations, namely $\alpha=10^{-4}$ and $10^{-3}$. The term on the right-hand side of Eq. \ref{eq:continuity} accounts for the formation of planetesimals. We will describe the details of our model for planetesimal formation later in this section (see Section \ref{sec:planetesimalformation}).

In Eq. \ref{eq:continuity}, a priori, we neglect the sublimation of ice dust grains  that drift inwards eventually crossing the disk iceline \citep[see in][]{morbidellietal16}. We will consider this effect later when modelling the growth of protoplanetary embryos.  We numerically solve Eq. \ref{eq:continuity} invoking the flux conserving implicit donor-cell algorithm with zero density boundary conditions \citep{birnstieletal12}. We use a radial logarithm-grid with 600 cells~\citep[][]{deschetal18}.

\subsection{Dust Coagulation model}

The degree of coupling between the gas and dust motions is well-defined by the so called Stokes number ($St$). Dust grains with $St\ll1$ are very well coupled to the gas flow streaming lines, whereas dust grains with $St\gg1$ are well detached from the gas motion. In the Epstein regime, the dust grain size is smaller than the mean free path of the gas and the Stokes number takes the following form \citep{epstein24}
\begin{equation}
    St = \frac{\pi a \rho}{2\Sigma_{\rm x}},
\end{equation}
where $a$ is the respective dust grain size, and $\rho$ is the particle bulk density.

The radial velocity of dust particles is computed as \citep[e.g.][]{okuzumietal12,uedaetal19}
\begin{equation}
v_{\rm r,dust} = -\frac{St}{St^2 +  (1 + Z')^2}2\eta v_{\rm k} + \frac{1+Z'}{St^2 + (1 + Z')^2 }v_{\rm r,gas}
\end{equation}
where $Z'$ represents the local gas-to-dust ratio.

In our simulations, the initial dust size is set equal to $a_0=10^{-4}~{\rm cm}$~\citep{birnstieletal12,drakzkowskaetal19}. These dust particles grow on the collisional timescale  to the following size 
\begin{equation}
a_{\rm grow}(t) = a_0\exp(t/\tau_{\rm grow}),
\end{equation}
where the growth timescale is defined as
\begin{equation}
\tau_{\rm grow} = \frac{\Sigma_{\rm x}}{\Sigma_{\rm dust} \Omega_{\rm k}}.
\end{equation}


The growth of dust particles is also controlled by drift and fragmentation. Overall, large dust grains tend to collide at higher relative velocities than small ones~\citep{zsometal10}. In the limit where the dust motion is dominated by the gas turbulent motion  one can define the largest size that dust grains can grow until they start to fragment due to high impact velocities. Following  \citep{birnstieletal12}, this maximum size ($a_{\rm frag}$) reads as
\begin{equation}
a_{\rm frag} = f_{\rm f}\frac{2 \Sigma_{\rm x}}{\pi \rho}\frac{v_{\rm f}^2}{3\alpha c_{\rm s}^2},    
\end{equation}
where ${\rm f_{\rm f}}~=~0.37$ is fudge parameter, and $v_{\rm f}$ represents the fragmentation threshold velocity. Based on the results of laboratory experiments \citep[e.g.][]{testietal14},  we assume that the fragmentation threshold velocities of silicate and ice dust grains are different~\citep{gundlachblum15}. We assume that silicate dust grains inside the snowline have  bulk densities equal to 3~${\rm g/cm^3}$ and  fragment at velocities of $1~{\rm m/s}$. Ice dust grains beyond the snowline -- which are expected to be more sticky -- have bulk densities equal to  1~${\rm g/cm^3}$ and fragments at velocities of $10~{\rm m/s}$. In our nominal experiments we model a smooth transition in the threshold fragmentation velocities and bulk densities at the snowline by invoking the following functions
\begin{equation}
v_{\rm f}(r) = \frac{10 + \exp{\left(\frac{r_{\rm snow}-r}{0.05r_{\rm snow}}\right)} }{ 1 + \exp{\left(\frac{r_{\rm snow}-r}{0.05r_{\rm snow}}\right)}}~~{\rm m/s}
\label{eq:vfrag}
\end{equation}

\begin{equation}
\rho(r)= \frac{1 + 3\exp{\left(\frac{r_{\rm snow}-r}{0.05r_{\rm snow}}\right)} }{ 1 + \exp{\left(\frac{r_{\rm snow}-r}{0.05r_{\rm snow}}\right)}}~~{\rm g/cm^3}.
\label{eq:bulk}
\end{equation}
In all simulations we set the $r_{\rm pb} = r_{\rm snow} $, where $r_{\rm snow}$ represents the location of the fixed disk water snowline in the disk. In our nominal simulations $r_{\rm snow}=5$~au, but we also performed simulations with a hotter disk where the snowline is at $\sim$10~au.

Dust fragmentation is also promoted by radial drift. The maximum dust grain size due to drift induced fragmentation is given as  \citep{birnstieletal12,drakzkowskaetal19}
\begin{equation}
a_{\rm df} =  2 f_{\rm f}  \frac{\Sigma_{\rm x}}{\rho \pi} \frac{v_{\rm f}}{\left|\eta\right| v_{\rm k}}.    
\end{equation}

Although fragmentation is the major limiting effect to dust sizes, this may not be case everywhere in the disk. Following \citep{drakzkowskaetal19}, we explicitly limit the size that dust particles grow in the disk before they drift. This effect is particularly important in the context of our two-dust specie  approach to model correctly dust evolution in the outer part of the disk, where the dust coagulation timescale can be longer than the drift timescale  \citep{drakzkowskaetal19}. The drift-limited dust size is 
\begin{equation}
a_{\rm drift} = \frac{f_{\rm d}}{\pi \rho} \frac{\Sigma_{\rm dust}}{2\left|\eta\right|},    
\end{equation}
where $f_{\rm d}~=~0.55$ is a fudge factor used in \citep{birnstieletal12} to better fit the results of dual dust-species model to those following the evolution of multiple  dust/pebble species. 

The largest pebbles in the disk have sizes given by
\begin{equation}
a_{\rm max} = \min(a_{\rm grow},a_{\rm frag},a_{\rm df},a_{\rm St =1},a_{\rm drift}),    
\end{equation}
where $a_{\rm St =1}$ represents the size of particles with $St=1$

Finally, we also impose that dust sizes can not get smaller than our initial dust-size
\begin{equation}
    a_{\max} = \max(a_{\rm max}, a_{\rm 0}).
\end{equation}

\subsection{Planetesimal Formation Model}\label{sec:planetesimalformation}

We will use the terms  ``dust'' and  ``pebbles'' to refer to our smallest and largest dust grains thereafter. As pebbles drift inwards due to gas drag they may be spontaneously  concentrated into clumps via vortices~\citep[e.g.][]{raettigetal15,lyraklahr11} or zonal flows~\cite[e.g.][]{johansenetal12,dittrichetal13,baistone14}. If the  local pebble  density reaches a critical threshold the clump may collapse to form planetesimals.  To mimic the formation of planetesimals in our simulations  we follow the approach of \cite{lenzetal19}. As shown in Eq. \ref{eq:continuity}, our analytic treatment of planetesimal formation appears as a sink term in the advection equation for dust/pebble evolution,  and it is given as
\begin{equation}
\frac{\partial \Sigma_{\rm pla}}{\partial t} =  - \frac{\varepsilon}{d}|v_{\rm r,peb}|\Sigma_{\rm peb}\mathcal{H},
\end{equation}
where 
\begin{equation}
\frac{\partial \Sigma_{\rm pla}}{\partial t} = \frac{\partial \Sigma_{\rm peb}}{\partial t}.
\end{equation}
$v_{\rm r,peb}$ and $\Sigma_{\rm peb}$ represent the radial velocity and surface density of pebbles in the disk, respectively. We set the pebble trap distance due to zonal flows or vortices as  $d=5H_{gas}$~\citep{lenzetal19}.  $\mathcal{H}$ is a heaviside step function that controls planetesimal formation \citep{gerbigetal19,lenzetal19}, and $\varepsilon$  is a dimensionless free parameter defining the planetesimal formation efficiency. In our simulations, we test several values for $\varepsilon$, from $10^{-3}$ to 0.8 in \citep{gerbigetal19,lenzetal19,voelkeletal20,gerbigetal20}. In this paper, we do not attempt to constrain $\varepsilon$. $\varepsilon$ may be constrained by using our final disks as input of simulations of the late stage of accretion of terrestrial planets and identifying those disks that better produce terrestrial planet analogues. A very massive terrestrial disk ($\lesssim2$~au) with total mass in solids larger than several Earth masses would probably fail to match the inner solar system terrestrial planets, in particular the low-mass Mars~\citep{chambers01,raymondetal05}. Meterorite ages is probably another constrain that can be used to bracket $\varepsilon$. Constraining plausible values of $\varepsilon$ will be valuable in any future studies.

Local planetesimal formation is only triggered if the following  condition is met 
\begin{equation}
 \dot{M}_{\rm peb,crit} \leq 2\pi r \Sigma_{\rm peb} |v_{\rm r,peb}|,
 \label{eq:criteria}
\end{equation}
where the critical pebble flux is
\begin{equation}
 \dot{M}_{\rm peb,crit} = \frac{m_{\rm c}}{\varepsilon\tau_{\rm c}}.  
  \label{eq:criteria2}
\end{equation}
The collapsing dust mass $m_{\rm c}$ is given by 
\begin{equation}
m_{\rm c } =\frac{4}{3} \pi l_{\rm c}^3 \rho_{\rm Hill}.
\label{eq:critflux}
\end{equation}
In Eq. \ref{eq:criteria2}, $\tau_{\rm c}$ represents the lifetime of traps that can concentrate pebbles to trigger planetesimal formation. We have set $\tau_{\rm c}=1000P_{\rm orb}$~\citep{baistone14}, where ${\rm P_{\rm orb}}$ is the orbital period at the clump-forming location ~\citep{lenzetal20}. This assumption is also consistent with particle concentration timescales observed, for instance, in numerical simulations of ``pure''\footnote{When pebble concentration occurs in the absence of zonal flows and vortices~\cite{klahretal18}.} streaming instability~\cite[e.g.][]{simonetal16,yangetal17}. In this work,  we assume that planetesimal formation occurs only  if $St>5\times10^{-5}$, which is a condition more strict than that used in \cite{voelkeletal20} ($St>0)$. Nevertheless, it may be considered too generous compared to the typical Stokes number required for planetesimal formation to take place in pure streaming instability~\cite[e.g.][]{yangetal17}. We will discuss the impact of this choice latter in the paper (in Section \ref{sec:discussion}). The  hill density~\citep{gerbigetal19} is defined as
\begin{equation}
    \rho_{\rm  Hill} = \frac{9}{4\pi}\frac{M_{\odot}}{r^3}
\end{equation}

By equating the diffusion and collapse timescale of the clump, ones derives the critical length scale \citep{lenzetal19}
\begin{equation}
l_{\rm c} = \frac{2}{3}\sqrt{\frac{\alpha}{St}}H_{\rm gas}.    
\end{equation}

Finally, if Eq. \ref{eq:criteria} is satisfied,  the surface density in planetesimals produced in a single timestep $dt$ is
\begin{equation}
  \Sigma_{\rm pla} =  \frac{\varepsilon}{d} |v_{\rm r,peb}|  \Sigma_{\rm peb}  dt
\end{equation}

\subsection{Planetary embryo's growth after streaming instability}

As planetesimals form in our simulations, we also model their growth via collisions with other planetesimals using a simple analytical prescription. We follow the growth of single planetesimals (thereafter we will refer to these bodies as protoplanetary embryos) at two specific locations of the disk: at 0.5 and 1.0~AU. Our goal here is not to build an entire system of embryos in the terrestrial region but instead roughly infer the typical masses of embryos in the terrestrial region at the end of the gas disk phase and their growth mode (via planetesimal or pebble accretion). Modeling the growth and evolution of several embryos during the late stage of accretion of terrestrial planets is beyond the scope of this paper.

The orbits of planetesimals in a gaseous disk are affected by several processes as gas drag damping, gravitational interaction with other planetesimals, viscous stirring and dynamical friction due to interactions with protoplanetary embryos, and collisional damping \cite[e.g.][]{idalin93,kokuboida00,thommesetal03,chambers06,morishimaetal13}. The interplay between these different processes  typically leads to an equilibrium state where the planetesimals' random velocities are in the dispersion dominated regime ($e_{pla},i_{pla}\gtrsim r_{\rm m H}/(2^{1/3}a_{\rm Emb})$), where
\begin{equation}
r_{\rm mH} \approx a_{\rm Emb}\left(\frac{2m_{\rm Emb}}{3M_{\odot}}\right)^{1/3}.
\end{equation}
$r_{\rm mH}$ represents the mutual hill radius of adjacent embryos with identical mass $m_{\rm Emb}$ and mean semi-major axis $a_{\rm Emb}$. 
In the limit where gravitational stirring and gas-drag damping are in equilibrium, the root-mean-square (rms) orbital eccentricities/inclinations of nearby planetesimals are defined as
\begin{equation}
e_{pla}=2i_{pla}=2.7\left( \frac{R_{\rm pla}\rho_{\rm pla}}{\Delta C_{\rm d}a_{\rm Emb} \rho_{\rm x}}  \right)^{1/5}\frac{r_{\rm mH}}{2^{1/3}a_{\rm Emb}},
\end{equation}
where $R_{\rm pla}$, $\rho_{\rm pla}$, $\Delta$, $\rho_{\rm x}$, $C_{\rm d}$ are the planetesimal radius, planetesimal bulk density, mutual separation of adjacent embryos in mutual hill radii~\citep{kokuboida00}, the gas column density at the embryo's position, and gas-drag coefficient. We have performed simulations with different sizes of planetesimals, $R_{\rm pla}$ varying from 10~Km to 500~Km. We assume a single planetesimal size in each simulation and neglect the effects of planetesimal fragmentation in our model~\citep[but see for instance][]{chambers06,chambers08}. The bulk density of planetesimals and embryos inside the snowline are set to 3~${\rm g/cm^3}$. Following~\cite{kokuboida00} we set $\Delta=10$, and $C_{\rm d}=1$~\citep{chambers06}.

\begin{figure}
\includegraphics[scale=0.35]{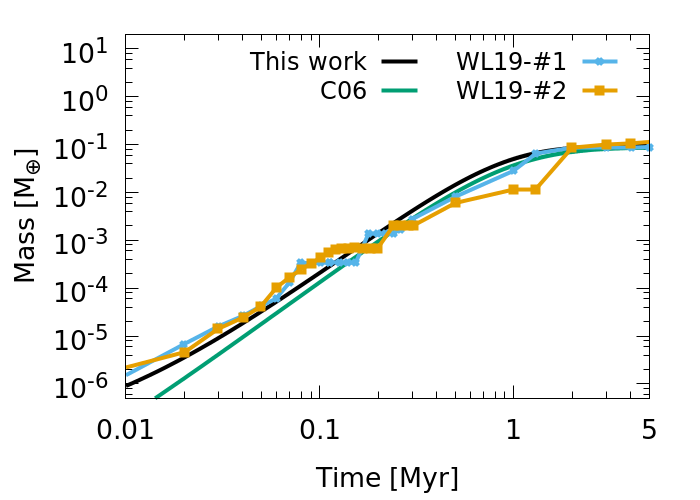}
\caption{Comparison of different planetesimal accretion prescriptions. In all scenarios, the growing embryo is at 1~au and assumed to not migrate. The underlying gasesous protoplanetary disk corresponds to a  MMSN disk \cite[e.g.][]{hayashi81}. Planetesimals are assumed to have sizes of 30~km for comparison purposes only. Our nominal simulations will assume different planetesimal sizes. The black line shows the growth rate of an embryo obtained considering our semi-analytical planetesimal accretion prescription. The green line shows the analytical prescription from \cite{chambers06}, and the blue and yellow lines show the growth obtained in high resolution N-body simulations \citep{walshlevison19}.}
\label{fig:fit}
\end{figure}

The growth of a protoplanetary embryo via accretion of planetesimals in the dispersion dominated regime is well described by the particle-in-a-box approach. The growth rate of an embryo depends on embryo's size, local planetesimal surface density, and on the relative velocities of  the embryo and planetesimals during close-approaches. The accretion rate of an embryo is given by \citep[e.g.][]{safronov1972,greenbergetal78,wetherill80,spauteetal91}
\begin{equation}
\frac{dM_{\rm Emb}}{dt}  \approx F \frac{\Sigma_{\rm pla}^*}{\sin(i_{\rm pla})a\sqrt{2}} \pi R_{\rm Emb}^2\left(1+\frac{v_{\rm esc}^2}{v_{\rm rel}^2}\right)v_{\rm rel},
\end{equation}
where $v_{esc}=\sqrt{\frac{2Gm_{\rm Emb}}{R_{\rm Emb}}}$ is the escape velocity at the embryo's surface, $v_{rel}$ the relative velocity between planetesimals and the planetary embryo. When calculating the planetary embryo's accretion rate, we update  $\Sigma_{pla}^*$ at every timestep to account for the reduction in the planetesimal surface density due to embryo-planetesimal accretion and also for the increase in surface density due to the formation of new planetesimals following our model. For simplicity, we assume that planetesimals and growing embryos have a negligible effect on the formation of new planetesimals. $F$ is a fudge factor invoked to compensate for the underestimated accretion rate when calculating the relative velocities between planetesimals and embryo using the RMS value of the planetesimals' random velocities. The deviation of planetesimals' velocity relative to a Keplerian circular and planar orbit (embryo's orbit) is defined as \citep[e.g.][]{kenyonluu98,morbidellietal09b}
\begin{equation}
v_{\rm rel}=v_{\rm k}\sqrt{\frac{5}{8}e_{\rm pla}^2+\frac{1}{2}(\sin{i_{\rm pla}})^2}.
\end{equation}
We have found that $F=3$ provides a good fit to the accretion rate of an embryo in N-body numerical simulations (see Figure \ref{fig:fit}; see also \cite{greenzweiglissauer92}). 

\subsection{Pebble accretion}
Pebble accretion onto protoplanetary embryos has been described in details in several studies~\citep[e.g.][]{ormelklahr10,lambrechtsjohansen12,johansenetal15,levisonetal15b,liuormel18,lambrechtsetal19}. Pebble accretion is well characterized by two regimes of accretion -- namely the Bondi and Hill accretion regimes. In the Bondi regime, the relative velocity between a pebble and a protoplanetary embryo on circular and coplanar orbit is
\begin{equation}
\Delta v = \frac{\sqrt{4St^2+1}}{St^2+1}\eta v_{\rm k}
\end{equation}
In the Hill regime, their relative velocity is higher and takes the form 
\begin{equation}
v_{\rm H}=\Omega_{\rm k}\frac{r_{\rm mH}}{2^{1/3}}. \end{equation}

A smooth transition from the Bondi to the Hill accretion regimes may be obtained by solving  the following combined velocity equation \citep{johansenetal15} using an interactive method
\begin{equation}
\hat{R}_{\rm acc} =  \sqrt{\frac{t_{\rm f} Gm_{\rm Emb}}{\delta v}}, 
\label{Eq:Racc}
\end{equation}
where $\delta v$ also depends on $\hat{R}_{\rm acc}$ as
\begin{equation}
\delta v(\hat{R}_{\rm acc}) = 0.25\Delta v + 0.25\Omega_{\rm k}\hat{R}_{\rm acc},    \end{equation}
and
\begin{equation}
t_{\rm f} = \frac{St}{\Omega_{\rm k}}.    
\end{equation}
 
We solve Eq. \ref{Eq:Racc} using the Newton-Raphson method for $\hat{R}_{\rm acc}$ by setting at first iteration $\delta v = \Delta v$.

The pebble surface density is calculated in Eq. \ref{eq:continuity} and the pebble surface density in the disk midplane is
\begin{equation}
\rho_{\rm peb,mid} = \frac{\Sigma_{\rm peb}}{H_{\rm peb}\sqrt{2\pi}}.    
\end{equation}
The pebble disk scale height is \citep[e.g.][]{johansenetal15}
\begin{equation}
    H_{\rm peb} = H\sqrt{\frac{\alpha}{St}}
\end{equation}

Finally, the pebble accretion rate of a protoplanetary embryo is
\begin{equation}
\frac{dM_{\rm Emb}}{dt} =4 \pi R_{\rm acc}^2 \rho_{\rm peb,mid}  \delta v (R_{\rm acc}),
\label{Eq:dMpebdt}
\end{equation}
where a good fit to the results of numerical integration's is obtained when 
\begin{equation}
R_{\rm acc}=\hat{R}_{\rm acc}\exp{(-0.4(t_{\rm f}/t_{\rm pass})^{0.65})}.
\label{Eq:Racc2}
\end{equation}
For a more detailed description of our pebble accretion prescription we refer the reader to previous works \citep{johansenetal15,lambrechtsetal19,izidoroetal19, bitschetal19}.
In Eq. \ref{Eq:Racc2},
\begin{equation}
t_{\rm pass} = \frac{GM_{\rm Emb}}{(\Delta v + v_{\rm H})^3}.
\end{equation}
Finally,  if $R_{\rm acc} < R_{\rm Emb}$ and $t_{\rm f} > t_{\rm  pass}$ we set $\frac{dM_{\rm Emb}}{dt} = 0$.

\section{Results} \label{sec:results}

\begin{figure*}
\centering
\includegraphics[scale=0.5]{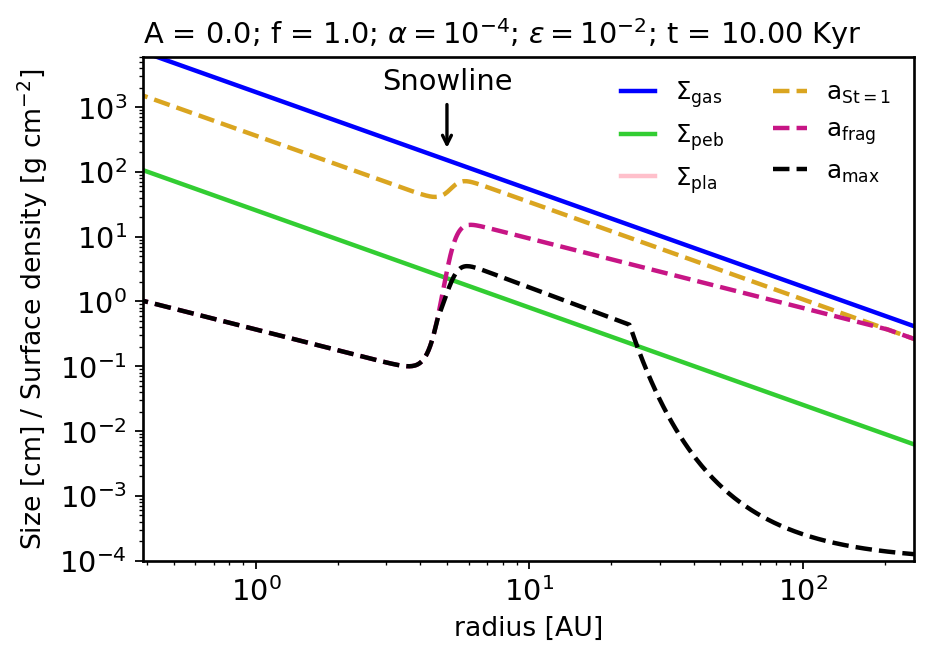}
\hspace{1.3cm}
\includegraphics[scale=0.5]{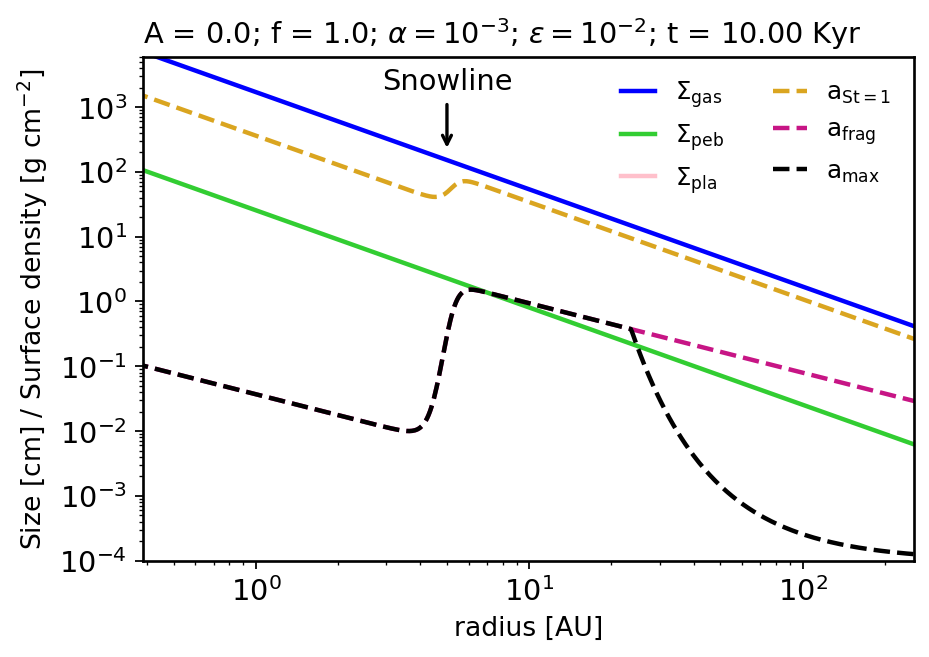}

\includegraphics[scale=0.5]{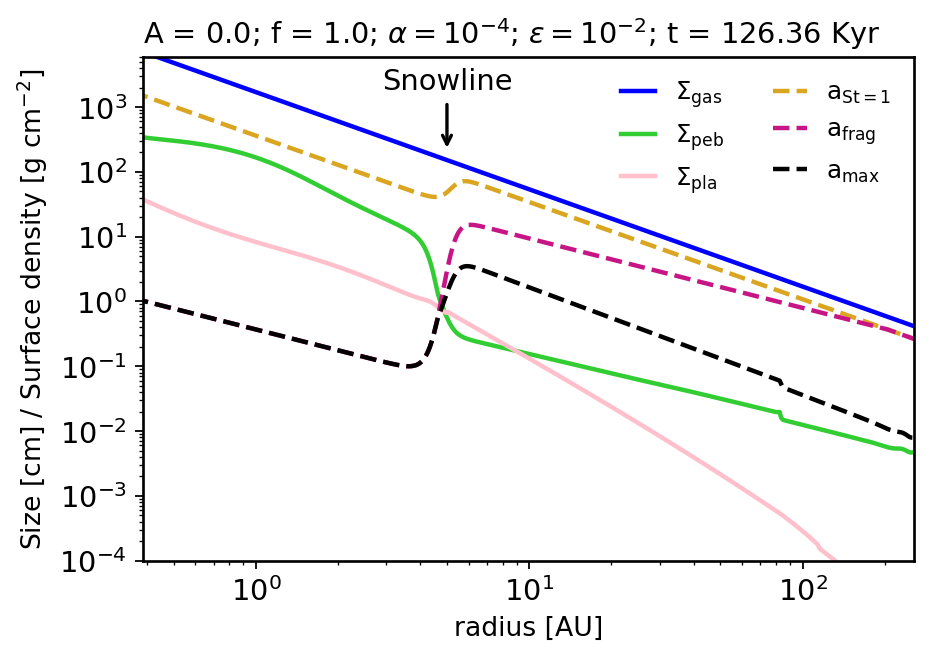}
\hspace{1.3cm}
\includegraphics[scale=0.5]{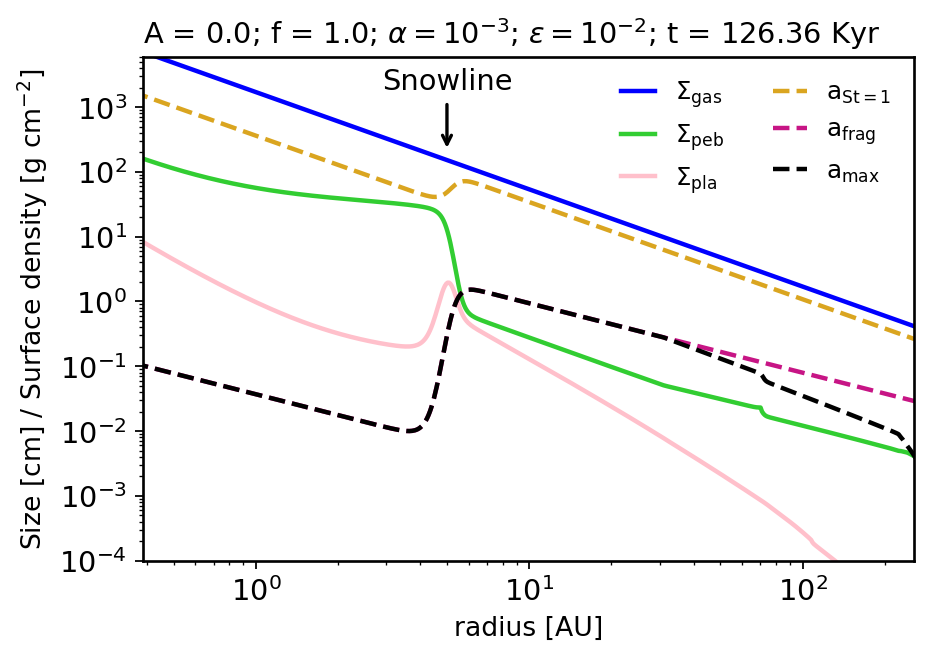}

\includegraphics[scale=0.5]{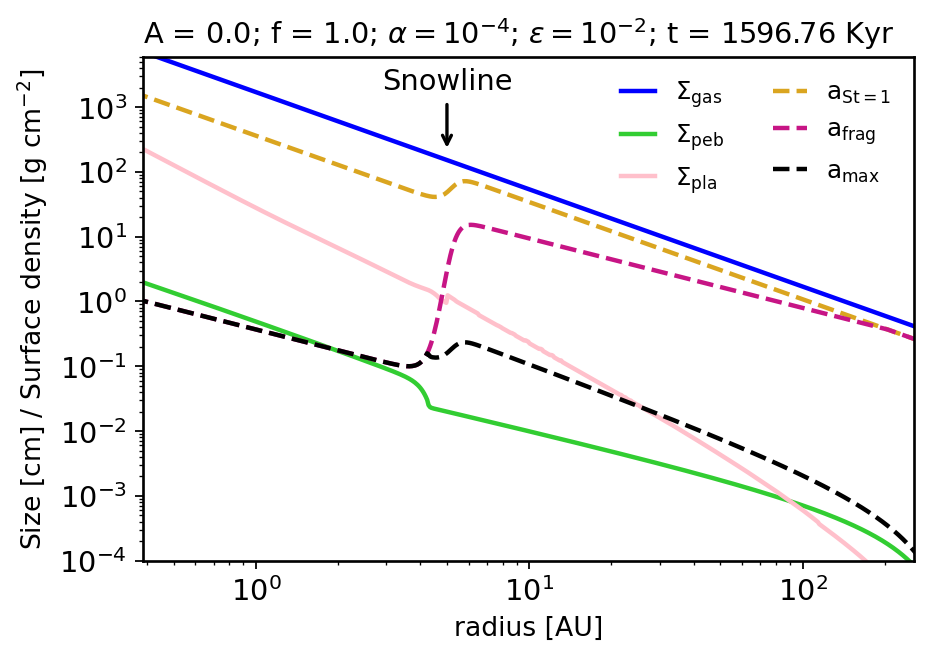}
\hspace{1.3cm}
\includegraphics[scale=0.5]{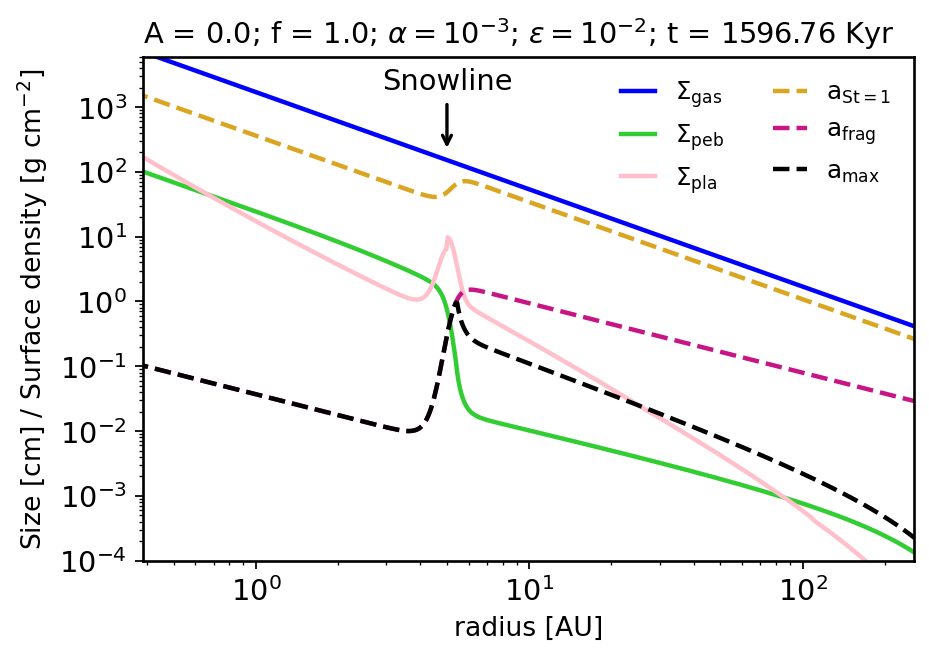}

\includegraphics[scale=0.5]{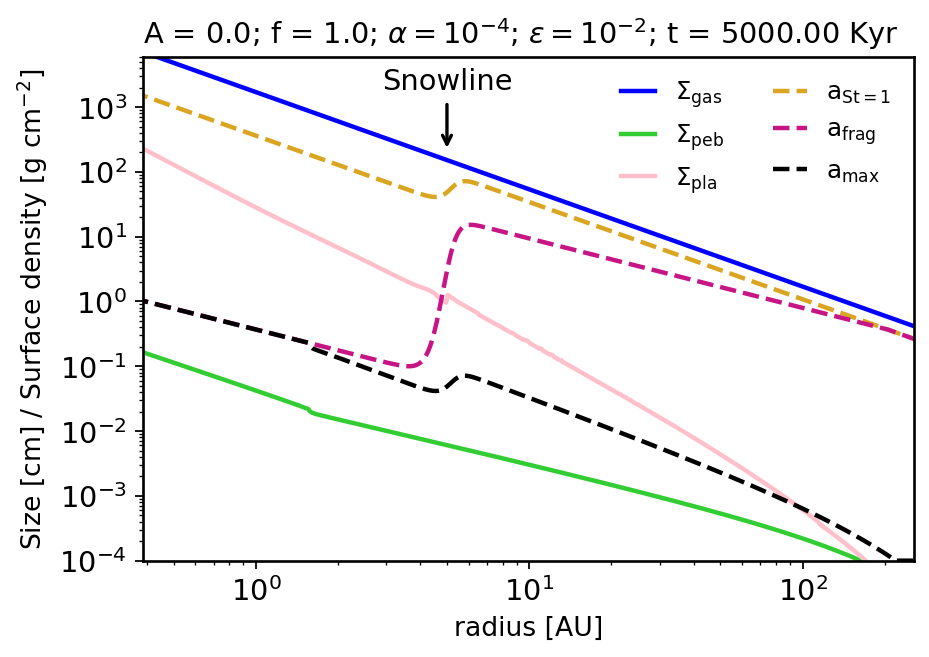}
\hspace{1.3cm}
\includegraphics[scale=0.5]{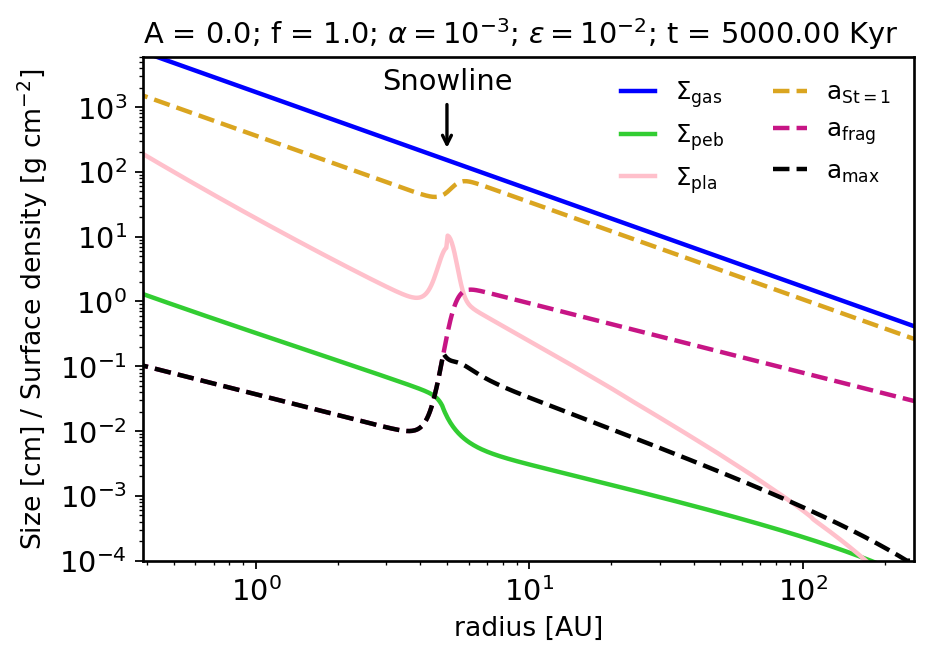}

    \caption{Snapshots of the evolution of pebble ($\Sigma_{\rm peb}$; green solid line) and planetesimal surface densities ($\Sigma_{\rm pla}$; light-pink solid line), and pebble sizes ($a_{\rm max}$; black dashed-line). The left and right panels corresponds to disks where $\alpha=10^{-4}$ and $\alpha=10^{-3}$, respectively. The time evolves from top to bottom. In both scenarios, the disk has no pressure bump ($A=0$). The gas surface density is kept constant over time (solid blue line). The maximum size that pebbles can reach before they start to fragment ($a_{\rm frag}$) is shown by the dashed violet-line. The sizes of particles with unity Stokes number is also shown for reference ($a_{\rm St=1}$; orange dashed line). In both simulations, the planetesimal formation efficiency is set $\epsilon=10^{-2}$.}
    \label{fig:surface_nobump}
\end{figure*}

Figure \ref{fig:surface_nobump} shows in a series of snapshots the evolution of pebble (green), planetesimal (light-pink), and gas surface (blue) densities in two of our simulations without pressure bump in the disk ($A$=0). On each panel we  also show the sizes of our pebbles ($a_{\rm max}$; dashed-black), fragmentation size ($a_{\rm f}$; violet), and particles sizes corresponding to $St=1$ ($a_{\rm St=1}$; orange). The time evolution of the system is shown from top to bottom. The water snowline is fixed at 5~au as indicated in every panel. The left-side and right-side panels show disks with $\alpha=10^{-4}$ and $\alpha=10^{-3}$, respectively. In both cases, the planetesimal formation efficiency is $\epsilon=10^{-2}$.

Figure \ref{fig:surface_nobump} shows that  pebbles in the disk with low-viscosity ($\alpha=10^{-4}$; left-side panels) grow larger than those in the disk with higher viscosity~\cite[e.g.][]{birnstieletal10,pinillaetal12}. The dips in the pebble sizes  (${\rm a_{max}}$; black) and ${\rm a_{frag}}$ (orange) at the snowline is due to the change in the threshold fragmentation velocity from 10~m/s to 1~m/s. The size of particles with $St=1$ also shows a dip because of the change in particles density at the snowline. From top to bottom, these panels show that as pebbles drift inwards, a fraction of them is converted into planetesimals. Pebbles  drift inwards and pile-up at about 5~au. This effect is observed because larger pebbles drift faster than smaller ones. When ice-pebbles reach the snowline, pebbles break into smaller silicate pebbles -- because of the change in the fragmentation velocity. These particles drift relatively slower, creating a pile up of pebbles (see also \cite{pinillaetal16}). The disk with $\alpha=10^{-3}$ shows a clear peak in the final surface density of planetesimal at the snowline. This is a consequence of the pebble traffic-jam at the snowline caused by the reduction in pebble size at this location~\citep{pinillaetal16,idaguillot16,drazkowskadullemond18}. A similar ``peak'' is  observed for the lower viscosity disk  but it rapidly disappears (before 0.1~Myr) because pebbles in the inner disk still drift relatively faster, due to their larger sizes. In our disk with higher viscosity the traffic-jam takes longer to dissolve so planetesimal formation is enhanced at this location.

\begin{figure*} 
\centering
\includegraphics[scale=0.5]{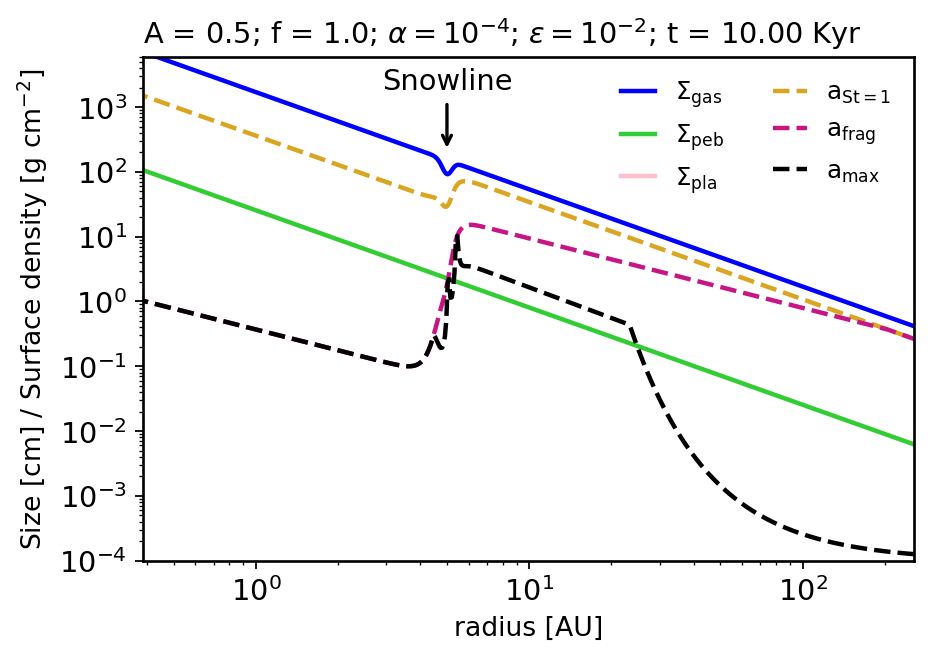}
\hspace{1.3cm}
\includegraphics[scale=0.5]{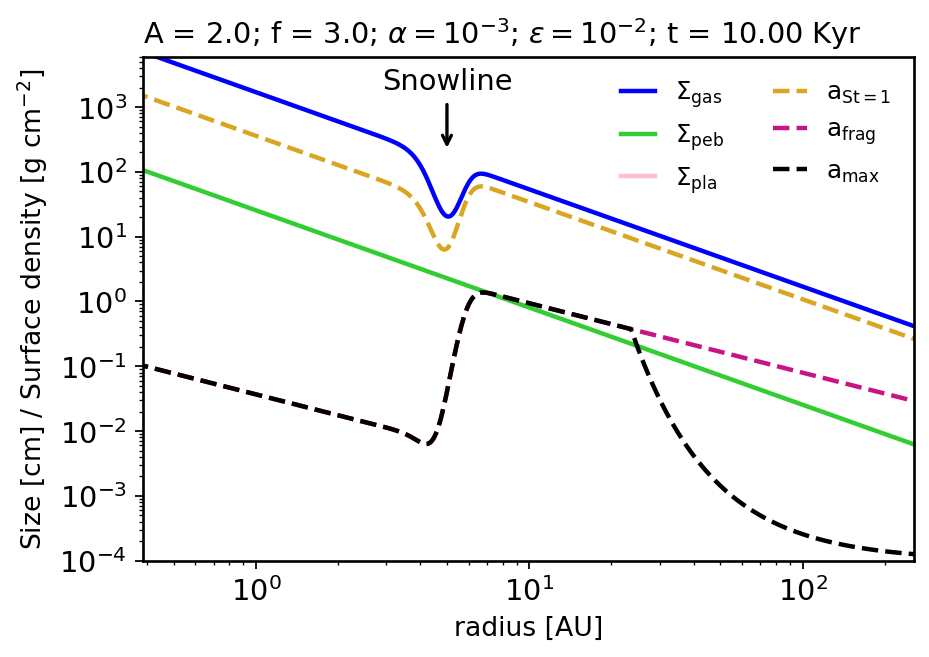}

\includegraphics[scale=0.5]{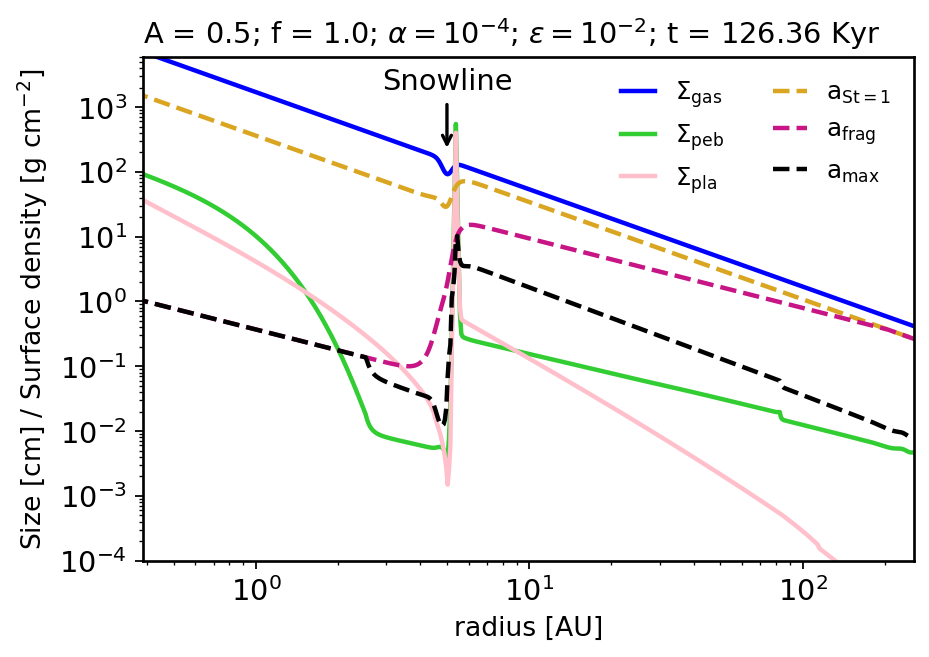}
\hspace{1.3cm}
\includegraphics[scale=0.5]{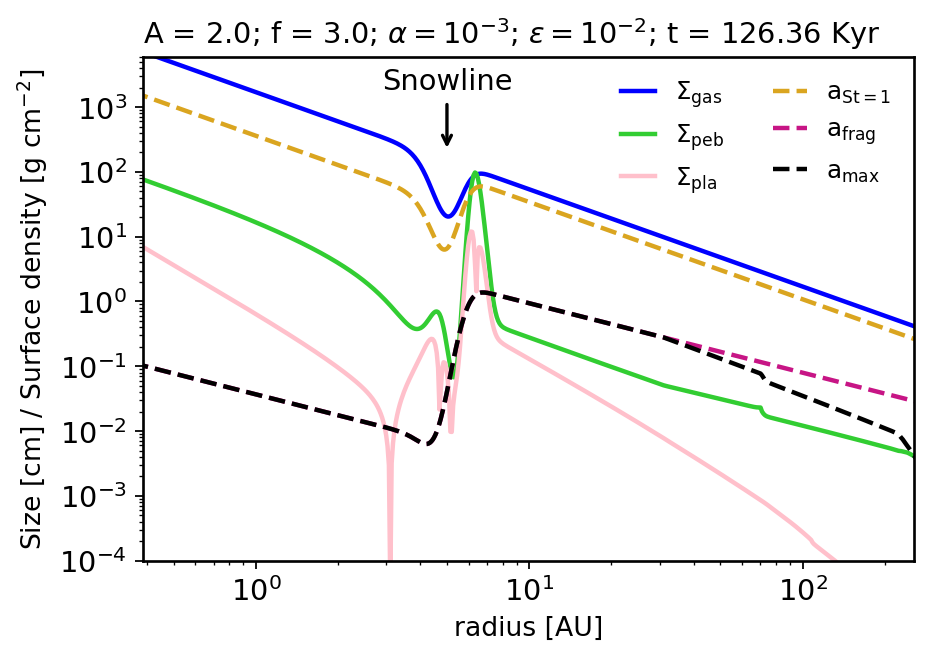}

\includegraphics[scale=0.5]{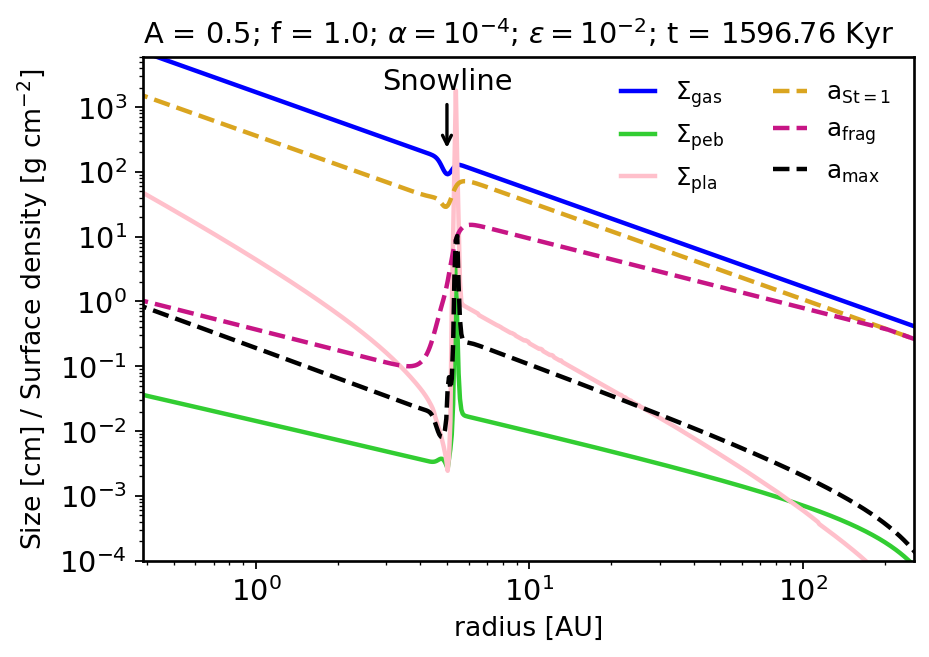}
\hspace{1.3cm}
\includegraphics[scale=0.5]{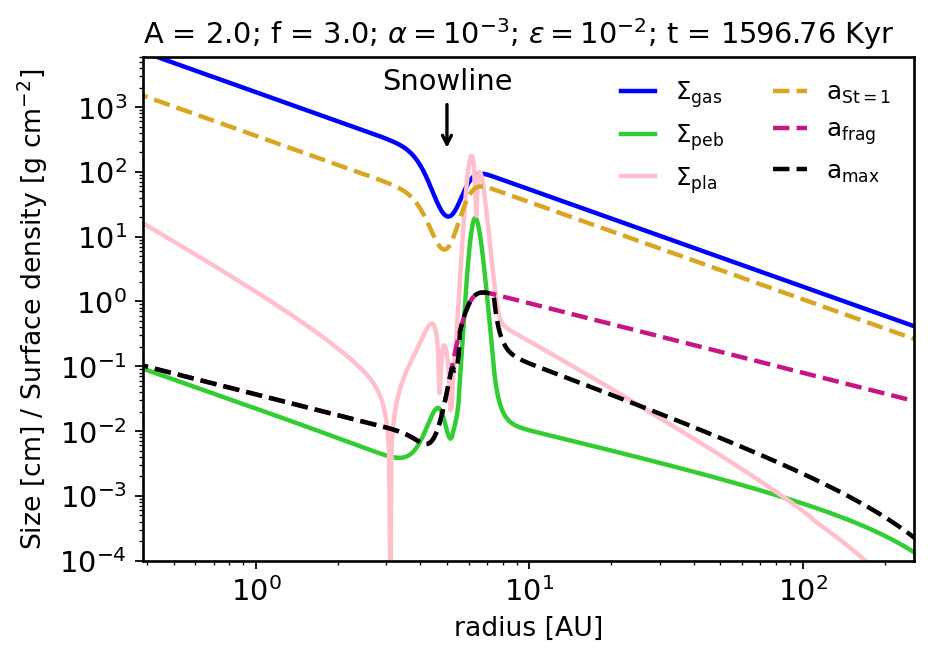}

\includegraphics[scale=0.5]{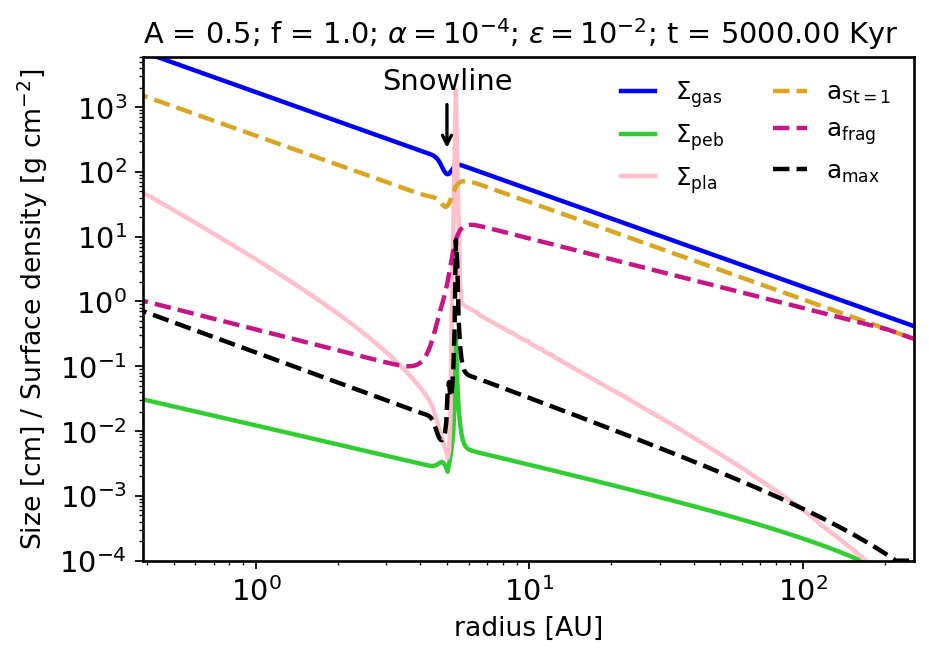}
\hspace{1.3cm}
\includegraphics[scale=0.5]{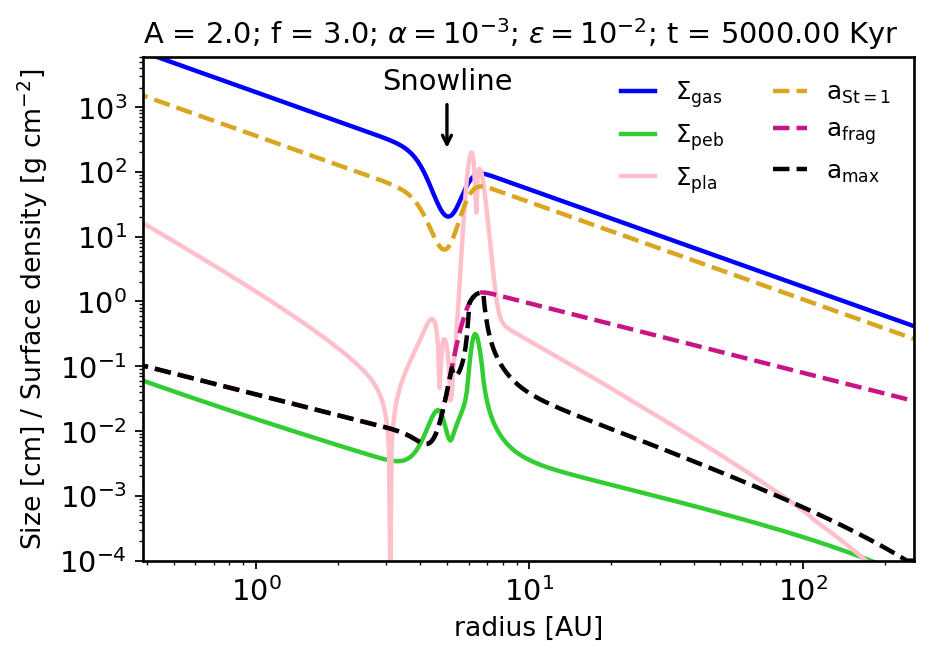}
    \caption{Snapshots of the evolution of pebble ($\Sigma_{\rm peb}$; green solid line) and planetesimal surface densities ($\Sigma_{\rm pla}$; light-pink solid line), and pebble sizes ($a_{\rm max}$; black dashed-line). \textbf{Left-panels:} Simulation with a pressure bump in a disk with $\alpha=10^{-4}$. The bump is assumed to exist in the disk since the beginning of our simulations, at 10~kyr. Pressure bump parameters are set $A=0.5$ and $f=1$. \textbf{Right:} Simulation with a pressure bump in a disk where  $\alpha=10^{-3}$. Pressure bump parameters are set $A=2.0$ and $f=3$. In both disk models, the bump parameters were chosen in order to provide a very efficient filtering of pebbles from the outer disk, where high-viscosity disks require deeper and wider bumps to achieve the same level of filtering. The gas surface density is kept constant over time (solid blue line). The maximum size that pebbles can reach before they start to fragment ($a_{\rm frag}$) is shown by dashed violet-line. The sizes of particles with unity Stokes number is also shown for reference ($a_{\rm St=1}$; orange dashed line). In both simulations, the planetesimal formation efficiency is set $\epsilon=10^{-2}$.}
    \label{fig:surface_withbump}
\end{figure*}

In our simulations with a pressure bump, the bump is either assumed to exist since the beginning of the simulation or appear at one specific time. We start by presenting the results of the first case.

Figure \ref{fig:surface_withbump} shows the results of two simulations where the disks starts with  pressure bumps. For each disk-viscosity, we have performed a series of test-simulations to identify the necessary pressure bump configurations -- namely width ($w$) and amplitude ($A$) --  that efficiently separate  the inner and outer solar system pebble/dust reservoirs once the bumps is fully formed. We will focus here on bump configurations where the pebble filtering provided by the bump is almost perfect. Note that the width and amplitude of the pressure bump that promote efficient filtering is different for different disk viscosities. We have found that in a disk where $\alpha=10^{-3}$, a bump with $A=2$ and $f=3$ provides very efficient filtering of pebbles. A lower-viscosity disk with $\alpha=10^{-4}$ show efficient filtering with a shallow bump,  where $A=0.5$ and $f=1$~\citep[see also][]{pinillaetal12}. A bump with $A=2$ is probably inconsistent with zonal flows caused by magneto-rotational instabilities~\citep{pinillaetal12} and it is probably more consistent with bumps caused by growing planets~\cite[e.g.][]{cridaetal06}. A bump with $A=0.5$, on the other hand, is roughly consistent with bumps seen in simulations of disks with zonal flows\footnote{Because \cite{pinillaetal12} invokes a cosine function rather a Gaussian function (compare our Eq. \ref{eq:1pb} and their Eq. 2) to mimic pressure bumps via gas density perturbation in the disks, a pressure bump with $A=0.5$ in our scenario roughly  corresponds to a bump with $A=0.25$ ($A/2$) in their model}~\citep{uribeetal11}. We have tested our code considering the pressure bump scenarios of ~\cite{pinillaetal12} and found good-agreement overall with our results.

\begin{figure*}
\centering
\includegraphics[scale=0.35]{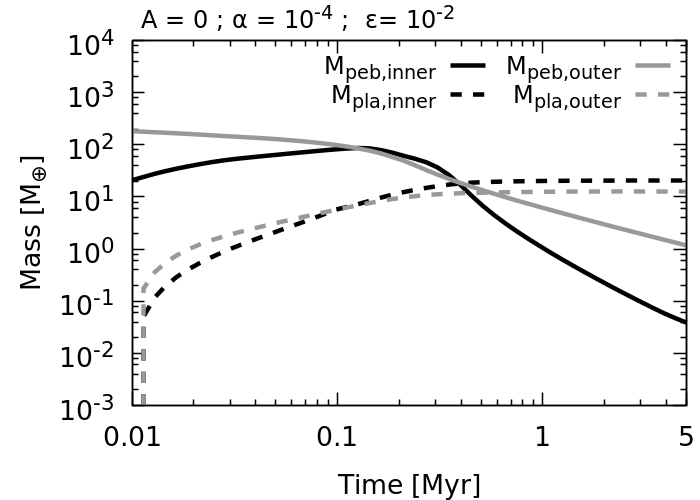}
\includegraphics[scale=0.35]{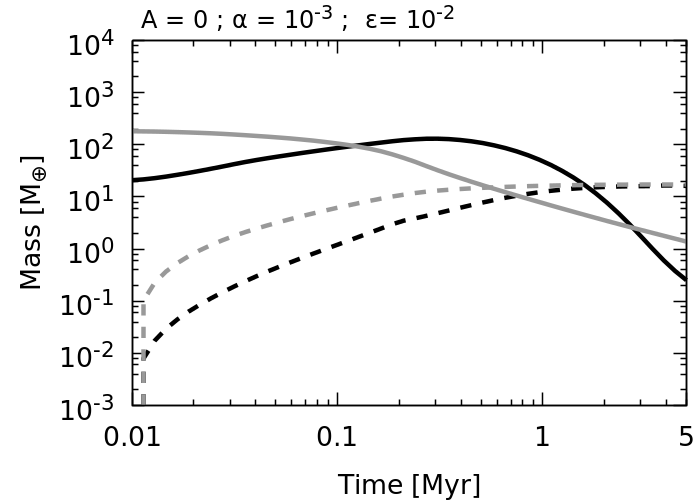}
\includegraphics[scale=0.35]{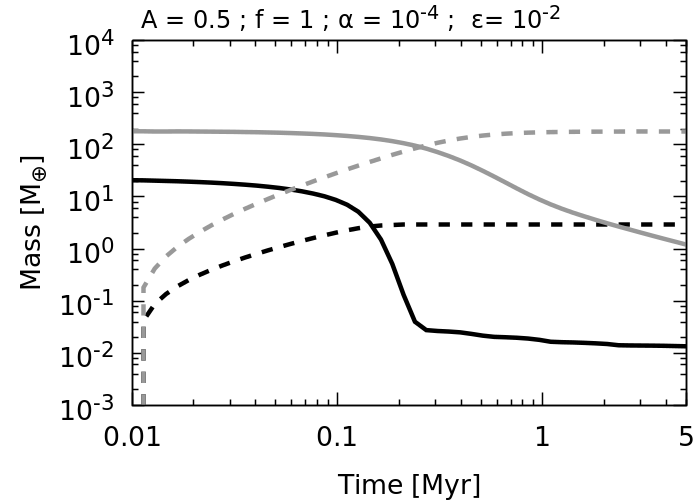}
\includegraphics[scale=0.35]{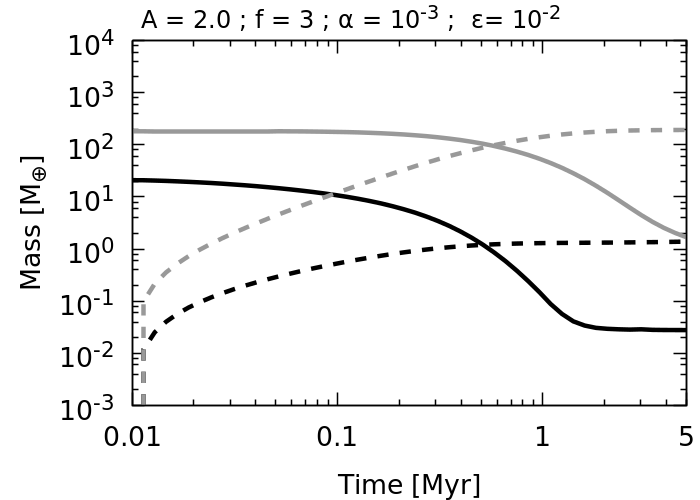}
\caption{Evolution of the total mass in different reservoirs of disks   without (top) and with pressure bumps (bottom). The total mass in pebbles (planetesimals) in the inner ($<$5~au) and outer ($>5$~au) parts of the disk are shown in solid (dashed) black and gray lines, respectively.  Top-panels show cases where the disk has no pressure bump (Figure \ref{fig:surface_nobump}). The bottom panels represent our nominal bump configurations (Figure \ref{fig:surface_withbump}). The disk viscosity and planetesiml formation efficiency are given at the top of each panel.}
\label{fig:diskmass}
\end{figure*}

In  Figure \ref{fig:surface_withbump}, the left and right-side panels show disks with different viscosities. In both cases, the inner ($<$5~au) and outer ($>$5~au) disks are very efficiently disconnected by the pressure bumps. n our nominal pressure bumps, the total pebble mass diffusing through the bump -- assuming that the ice component sublimates when icy-pebbles cross the snowline and it accounts for 50\% of the pebble mass --
is always less than $\sim$0.5~${\rm M_{\oplus}}$ ($M_{\oplus}$ represents 1 Earth mass), and in some cases as low as  $\sim$0.05~${\rm M_{\oplus}}$. Since only a fraction of the this dust is converted into planetesimals, our inner and outer disk remain compositionally distinct if one assumes that they start with different (isotopic) compositions. Note that stronger pressure bumps (wider and deeper) or bump in disks with lower turbulent viscosities would be even more efficient in disconnecting the inner and outer reservoirs without affecting our main results. Pebbles beyond the bump pile-up and form planetesimals whereas pebbles in the inner disk are free to drift inwards. Consequently, the inner disk is rapidly depleted of pebbles. In the disk with low-viscosity - where pebbles are relatively larger - pebbles are lost in about 0.2~Myr (after this time, embryos growth by pebble accretion is marginal, due to residual pebbles in the disk). In the high-viscosity disk, pebbles in the inner disk drift relatively slowly and the inner disk loses most of its mass in pebbles in about 1~Myr. The disk mass evolution in pebbles and planetesimals in the inner and outer disk in simulations with and without pressure bumps (Figues \ref{fig:surface_nobump} and \ref{fig:surface_withbump}) are shown in Figure \ref{fig:diskmass}.
 
 In simulations with no-pressure bump, the disk lose most of the pebbles quickly. In each of these scenarios, the total mass in planetesimals in the inner and outer disk at the end of simulation is comparable, within a factor of a few. Of course this result  is a direct consequence of our boundary conditions, where at the disk inner edge all pebbles are lost. At $\sim$1~Myr, the total pebble mass beyond 5~au in our disks without pressure bumps is about $\sim~6{\rm M_{\oplus}}$ for $\alpha=10^{-4}$ and  $\sim~8M_{\oplus}$ for $\alpha=10^{-3}$. At 3~Myr, these quantities drop to $\sim~2{\rm M_{\oplus}}$ for $\alpha=10^{-4}$ and  $\sim~2.3{\rm M_{\oplus}}$ for $\alpha=10^{-3}$. Note that the total mass in pebbles  also diminish with time due to planetesimal formation in the disk. The estimated total dust/pebble mass in 1 to 3~Myr old observed disks around stars with masses around 0.5-1.5$M_{\odot}$ ($M_{\odot}$ represents 1 solar mass) varies between $\sim$0.3${\rm M_{\oplus}}$ and $\sim$160${\rm M_{\oplus}}$~\cite[e.g.][]{manaraetal18}. Therefore, the typical pebble mass available in our disks at 1 and 3~Myr is at least marginally consistent with observations, in particular with Class II disks, where the median dust mass is $\sim$3~${\rm M_{\oplus}}$ ~\citep{tychoniecetal20}. Class-0 and Class-I disks are much more massive than Class-II disks~\citep{tychoniecetal20}. Planetesimal-planetesimal collisions may also play a role in replenishing the disk with second generation dust~\cite[e.g.][]{gerbigetal19}. This effect is not included in our model. Observed disks typically exhibit  ring-like structures in the dust distribution,  which suggest that pressure bumps have concentrated dust and pebbles at specific locations of the disk~\cite[e.g.][]{dullemondetal18}. However, the fraction of the initial  dust/pebble mass that has been already used for planet formation or lost via gas-drag drift in these disks is not constrained by observations. In our simulations where the pressure bump is present since the beginning, most  pebbles in the outer disk drift inwards and pile up at the pressure bump. As shown in Figure \ref{fig:surface_withbump}, this resulting pile of pebble  leads to very efficient planetesimal formation at this location (see peak shown in the pink-line at 5~Myr in the bottom panels of Figure \ref{fig:surface_withbump}). Although planetesimal formation is expected to be quite efficient in the bump \citep{carreraetal20}, planetesimals should collide with each other and eventually grow to (multiple) planetary embryos which probably self-regulate planetesimal formation in this region~\cite[e.g.][]{drazkowskadullemond14}. We do not include this effect in our simulations because our main goal here is not to model planetary growth in the bump. A giant planet core would probably grow very rapidly in our bump~\citep{morbidelli20,guilleraetal20}. Thus, the final total mass in planetesimals in the bump location in all our simulations is  probably overestimated. In our simulations with pressure bumps, the total mass in planetesimals in the inner disk  is between $\sim$~1 and $\sim~4M_{\oplus}$. Of course, these numbers may largely change depending on the planetesimal formation efficiency assumed.

%


\subsection{Planetary embryo's growth from planetesimal and pebble accretion}

To model the growth of planetary embryos from planetesimal and pebble accretion, we use the output of our dust evolution simulations, viz. the evolution of planetesimal surface density and pebble flux in the inner disk. 
 Planetary embryos in our simuations grow from  planetesimal masses. The initial nominal radius of our protoplanetary embryos is 100~Km\footnote{This also corresponds to the typical sizes of planetesimals forming in streaming instability simulations~\citep{johansenetal15,simonetal16} and this size is consistent with models of the asteroid belt evolution~\citep{morbidellietal09b}}, which corresponds to a mass of about 3$\times10^{-6}{\rm M_{\oplus}}$ for spherical planetesimal with uniform  density of 3${\rm g/cm^2}$ . All individual planetesimals in the feeding zone of a growing embryo are assumed to have the initial radius and mass of the embryo and to not grow in time. As mentioned before, we will perform simulations assuming different initial radii for planetary embryos (see Section \ref{sec:sizes}), ranging from 10 to 500~km to test how the initial planetesimal/embryo size impacts our results.

We do not model the growth of planetary embryos in the entire inner disk.
We estimate the final masses and the growth mode of planetary embryos at 0.5~au and 1~au only. This approach is sufficient to provide  information about the typical masses and growth mode of embryos in the terrestrial zone. During embryos' growth, we neglect the effects of type-I migration. We will discuss how it may impact our results later in the paper. In this section, we restrict ourselves to simulations with pressure bumps in the disk. Without pressure bumps planetary embryos in the inner disk are expected to grow to masses beyond one Earth mass~\citep{lambrechtsetal19,voelkeletal20}.

\begin{figure*}
    \centering
    \includegraphics[scale=0.35]{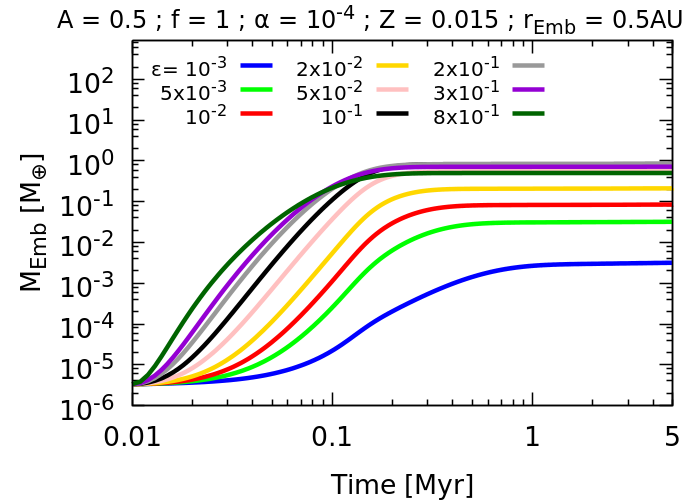}
    \includegraphics[scale=0.35]{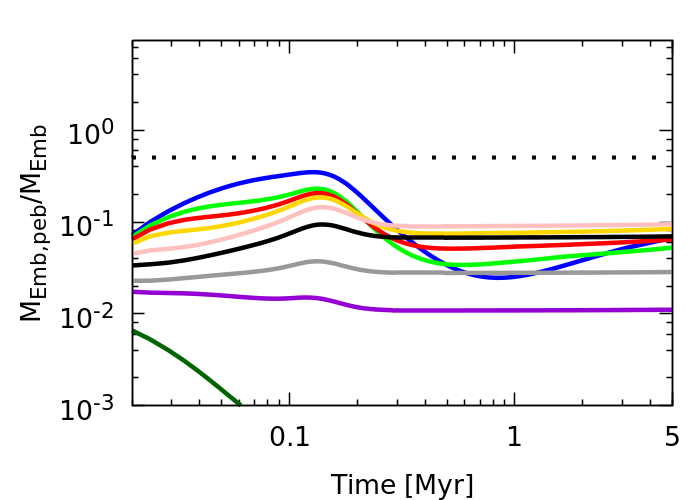}
    \includegraphics[scale=0.35]{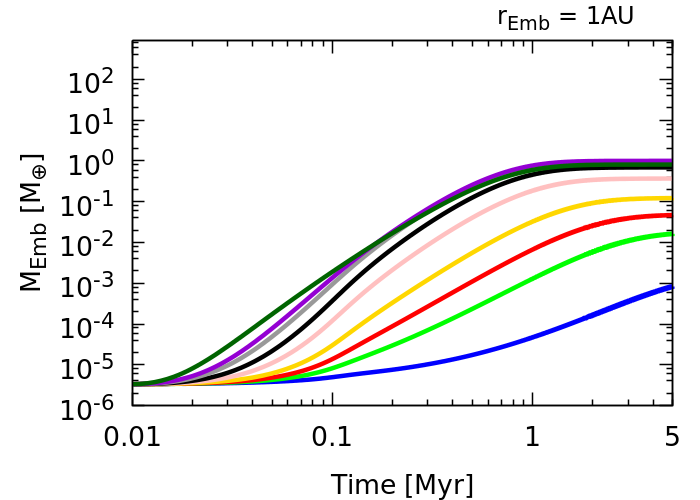}
    \includegraphics[scale=0.35]{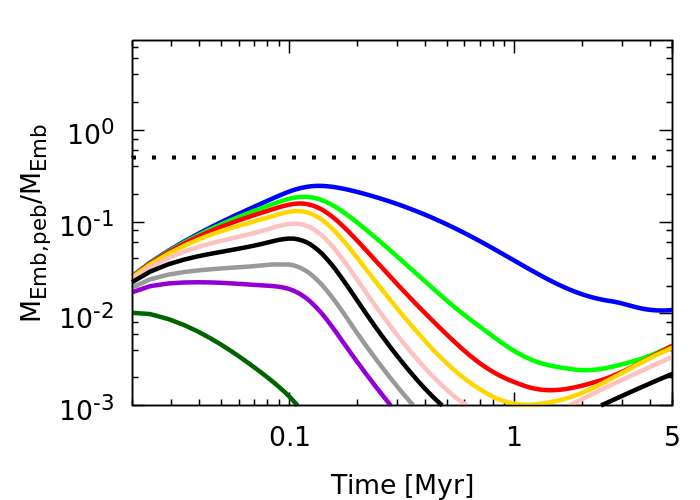}
    \caption{\textbf{Left:} growth of planetary embryos in simulations with different planetesimal formation efficiencies ($\epsilon$, varying from $10^{-3}$ to $0.8$). \textbf{Right} Fractional contribution of pebble accretion to the embryo's mass. The horizontal dashed line represents 50\% contribution (half of the embryo's mass comes via pebble accretion and the other half from planetesimal accretion). The top panels show embryos growing at 0.5~au. The bottom panel show embryos growing at 1~au. Both embryos are considered on non-migrating orbits. The different colors show individual embryos growing in disks with different $\epsilon$ (planetesimal formation efficiency). This case corresponds to a disk with $\alpha=10^{-4}$. The bump parameters are on the top-left panel (see Figure \ref{fig:surface_withbump}; left-side panels).}
    \label{fig:growthlowvisc}
\end{figure*}

Figure \ref{fig:growthlowvisc} shows the growth of planetesimals in disks with $\alpha=10^{-4}$ and a pressure bump (see bump parameters on the left-top panel)  with different planetesimal formation efficiencies.
The left-panels show embryo's mass evolution as a function of time. The right-side panels show the relative contribution of pebbles (via pebble accretion) to the final embryo's  mass. The top-panels of Figure \ref{fig:growthlowvisc} correspond to an embryo growing at 0.5~au whereas the bottom panels correspond to an embryo growing at 1~au. It can be seen that the final masses of the embryos may vary up to 3 orders of magnitude, depending on the planetesimal formation efficiency.  However, embryos at both locations grow systematically from accreting other planetesimals rather than pebbles. In fact the contribution of pebble accretion to embryos growing at 1~au is even lower, compared to an embryo at 0.5~au. This result does not depend much on planetesimal formation efficiency, but is instead caused by the fast inward drift of pebbles in the inner disk. Embryos in the innermost parts of the disk grow at a faster rate by first accreting other planetesimals - because of shorter dynamical timescales - than an embryo at 1~au. The embryo at 1au takes longer to reach masses where pebble accretion becomes relatively more efficient and pebbles in the inner disk are lost before it can accrete them. This result is a norm of almost all our simulations. Growing planetary embryos eventually stop growing via planetesimal accretion because as they grow they increase the random velocities and reduce the density of planetesimals in their feeding zones. Figure \ref{fig:growthlowvisc} shows that the growth timescale of Earth-mass embryos is significantly shorter than the Earth's accretion timescale, which is estimated to be between 30-150~Myr after the solar system formation ~\citep{woodhalliday05,jacobsen05,kleinewalker17}. In some scenarios, even Mars-mass embryos grow in timescales shorter than that of Mars' which is estimated to be about 1-10~Myr~\citep{dauphaspourmand11}. This is particularly true in our disks with high planetesimal formation efficiency. Our planetesimal-planetesimal accretion prescription, which is based on the oligarchic growth regime, remains particularly valid for planetary embryos with masses of about Earth-mass or lower. Very massive planetesimal disks rapidly produce very massive planetary embryos -- in particular in the inner regions ($<$1au). Disks with massive planetary embryos transition  into the subsequent chaotic growth regime earlier than disks with lower-mass embryos~\citep{kokuboida02,kenyonbromley06,walshlevison19}. Therefore, in reality, in the late stages of our disks, planetary embryos in the inner parts of the disk could be even more massive than our final estimates.

\begin{figure*}
    \centering
    \includegraphics[scale=0.35]{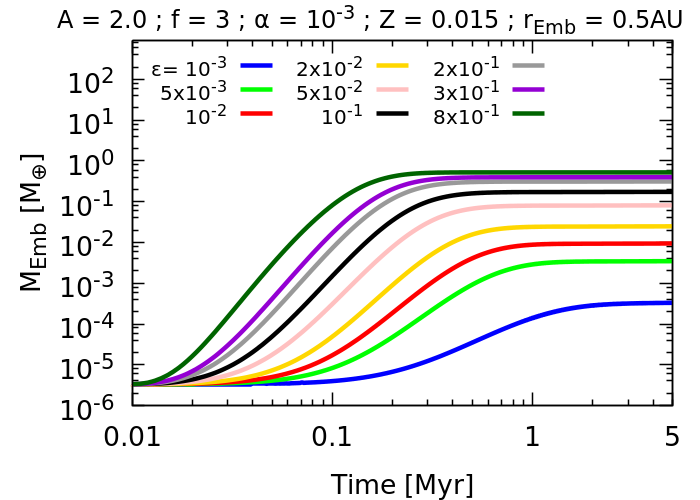}
    \includegraphics[scale=0.35]{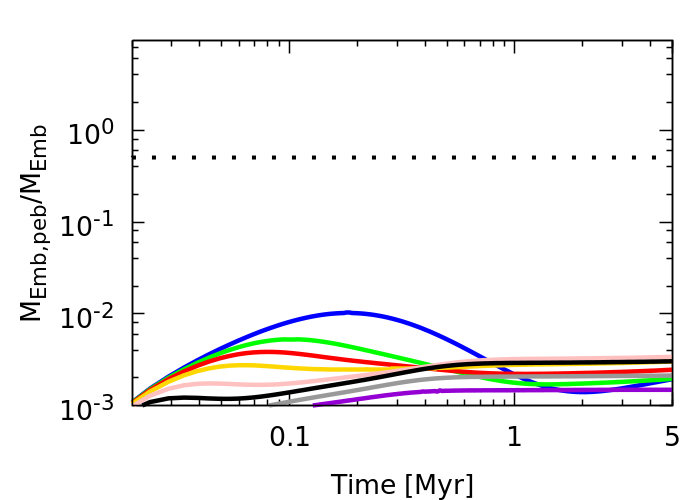}
    \includegraphics[scale=0.35]{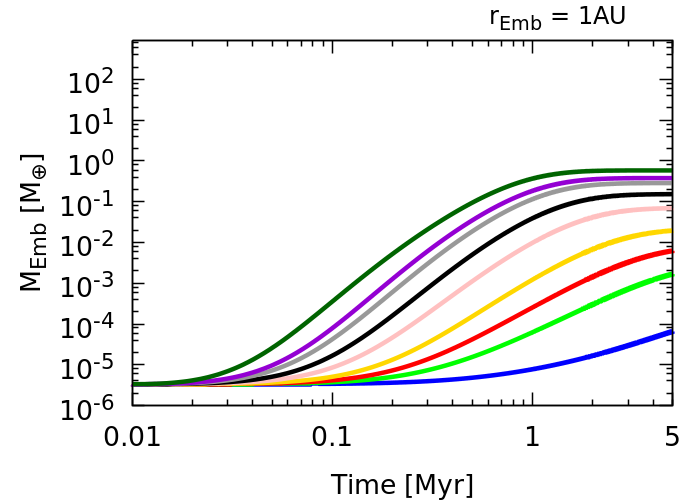}
    \includegraphics[scale=0.35]{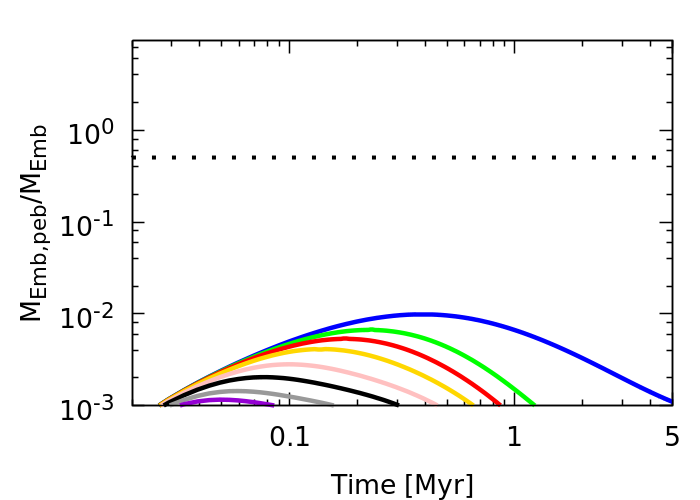}
    \caption{\textbf{Left:} growth  of planetary embryos in simulations with different planetesimal formation efficiencies ($\epsilon$, varying from $10^{-3}$ to $0.8$). \textbf{Right} Fractional contribution of pebble accretion to the embryo's mass.
    The horizontal dashed line represents 50\% contribution (half of the embryo's mass comes via pebble accretion and the other half from planetesimal accretion)The top panels show embryos growing at 0.5~au. The bottom panel show embryos growing at 1~au. The different color show individual embryos growing in disks with different $\epsilon$. This case corresponds to a disk with $\alpha=10^{-3}$. The bump parameters are shown on the top-left panel (see Figure \ref{fig:surface_withbump}; right-side panels).}
    \label{fig:growthhighvisc}
\end{figure*}

Figure \ref{fig:growthhighvisc} shows the evolution of embryos in our high-viscosity disk. In this scenario, because pebbles in the inner disk are even smaller than those in the low-viscosity disks, they are even less efficiently accreted by growing embryos. Therefore, in this case, the contribution of pebble accretion to the growth of embryos in the terrestrial region is less than 1\%.

\subsubsection{Dependence on the initial planetesimal size}\label{sec:sizes}

\begin{figure*}
    \centering
    \includegraphics[scale=0.35]{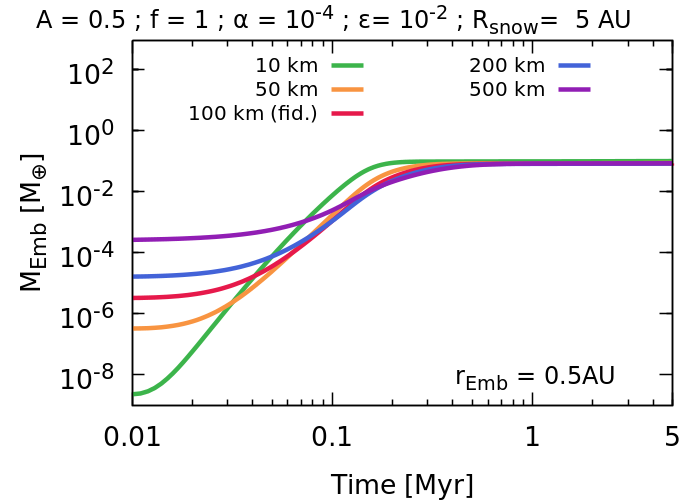}
    \includegraphics[scale=0.35]{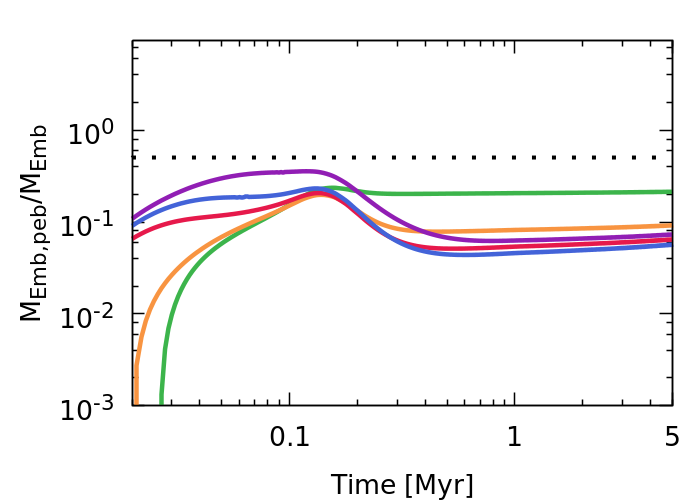}
    \includegraphics[scale=0.35]{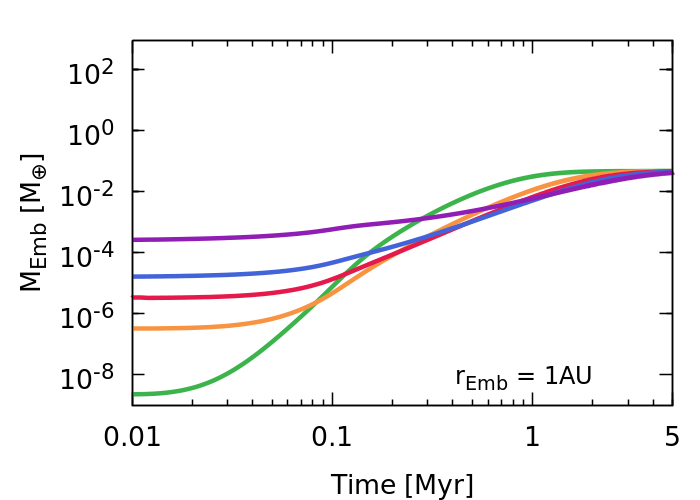}
    \includegraphics[scale=0.35]{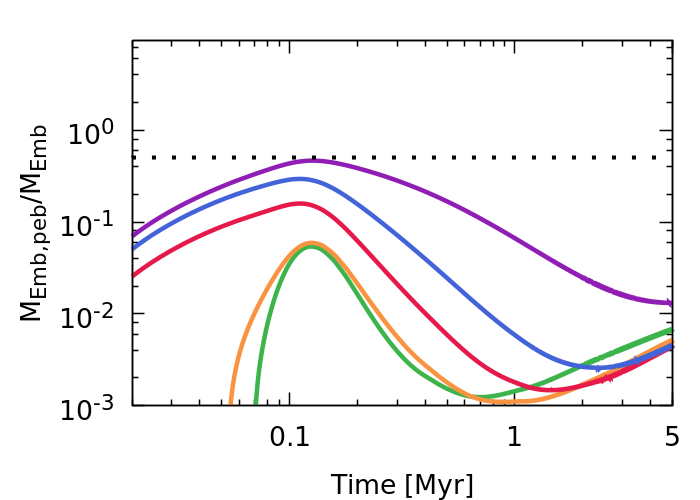}
    \caption{Growth of planetary embryos in simulations with different  planetesimal radii. The different colors show planetesimal with different radii as indicated on the top-left panel. We note that in each simulation the initial mass of the planetary embryo correspond to that of the individual planetesimals. The right-side panels show the contribution of pebble accretion to the embry's mass. The top panels show embryos growing at 0.5~au. The bottom panel show embryos growing at 1~au. This case corresponds to a disk with $\alpha=10^{-4}$. The bump parameters are shown on the top-left panel (see Figure \ref{fig:surface_withbump}; left-side panels).}
 \label{fig:lowvisc_diffsize}
\end{figure*}

Streaming instability simulations produce a distribution of planetesimal sizes, with the smallest ones having radii of about $50$km to the largest ones with radii of  $\sim$500km~\citep[e.g.][]{johansenetal15,simonetal16}. We have performed a set of simulations to test the dependence of our results on planetesimal sizes. Figure \ref{fig:lowvisc_diffsize} shows the growth of an embryo at 0.5~au (top-panels) and 1~au (bottom-panels) in disks with different initial planetesimal sizes. Our planetesimals' radii vary from 10~km to 500~km. We have found that  disks with smaller planetesimals promote faster growth~\cite[e.g.][]{levisonetal10} because planetesimals accrete more efficiently from the beginning. However, even in this faster growth regime, the contribution of pebble accretion to the final embryo mass is always lower than 50\%. This is also the case if we assume that 500~km planetesimals grow in a sea of 10~km planetesimals. Even if the embryos grow faster via accretion of small planetesimals, pebbles in the inner disk are lost before  growing embryos can gain significant mass via pebble accretion. Again, the contribution of pebble accretion to embryos growing at 1~au, is relatively smaller. We have also performed simulations for a disk with $\alpha=10^{-3}$ and the results were qualitatively similar.

\subsubsection{Effects of the gas surface density}\label{sec:effectgas}

\begin{figure*}
    \centering
    \includegraphics[scale=0.35]{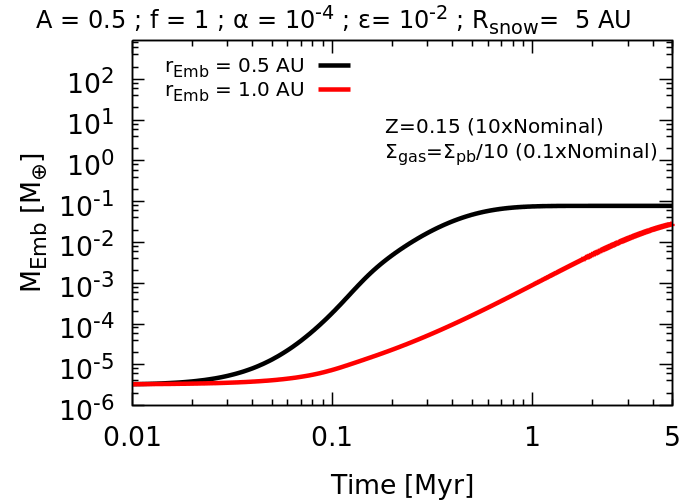}
    \includegraphics[scale=0.35]{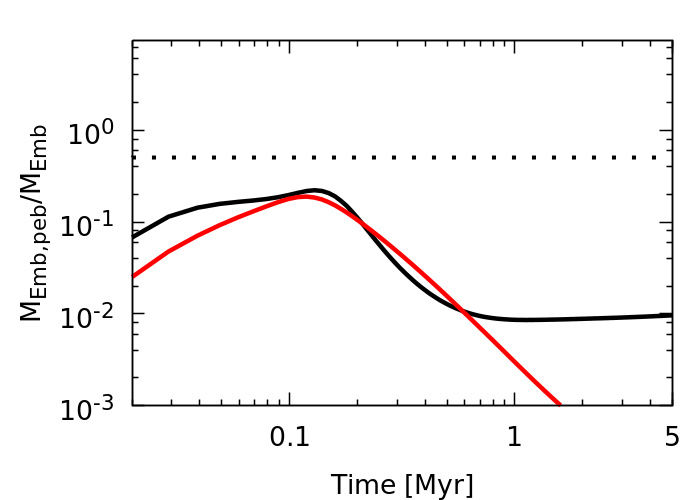}
    \caption{Growth of planetary embryos in simulations starting with a low-mass gaseous disk and high dust-to-gas ratio. The black line shows an embryo growing at 0.5~au whereas the red line represents an embryo growing at 1~au. Planetesimal size in this case corresponds to a radius of 100~Km.}
    \label{fig:embryosuperhighZ}
\end{figure*}

In our simulations, for simplicity, we assume that the disk is not evolving with time because we are mostly interested in the early stages of the disk evolution. However, in order to understand how the gas surface density may impact our results, we performed an additional set of simulations where we reduce the initial gas surface density by a factor of 10 ($\Sigma_{\rm mmsn}/10$). We keep the assumption that the gas surface density does not evolve in time. In our simulations invoking a  low-mass disk scenario, we increase the initial disk dust-to-gas ratio by a factor of 10 in order to have initially the same amount of dust in the disk compared to our nominal disk. This increase in the dust-to-gas ratio may also facilitate planetesimal formation via streaming instability~\citep{simonetal16,yangetal17}. Figure \ref{fig:embryosuperhighZ} shows that embryos growing both at 0.5~au and 1~au in a low-mass disk also grow mostly via planetsimal accretion. In Figure \ref{fig:embryosuperhighZ},  planetesimal radius (and initial embryos size) is 100~km.

\subsubsection{Simulations with growing pressure bumps}

So far in our model set up, the pressure bump in the disk was either absent or was assumed to exist since the beginning of the simulation. Here we test the effects of the pressure bump appearing at later times of the disk. In this case, pebbles from the outer disk drift into the inner solar system until the pressure bump forms. We assume that these pebbles also have NC-like isotopic compositions. If these pebbles have CC-like compositions that would perhaps violate the solar system isotopic dichotomy~\citep[but see][]{schilleretal18,Schilleretal19}. In these simulations, we reduce the pebble flux by a factor of 2 at the snowline when computing planetary embryo's growth.

In order to mimic the growth of the pressure bump with time, we use linear interpolation between our power-law gas disk profile ($\Sigma_{\rm mmsn}$) and nominal bumpy-disk ($\Sigma_{\rm pb}$). We test three scenarios. In the first one, the bump linearly grows from 0.1~Myr to 0.2~Myr. In the second case, it grows from 0.2~Myr to 0.3~Myr. In both scenarios the bump is completely formed before 0.5~Myr, otherwise most of the pebbles from the outer disk reservoir drift inwards potentially making it difficult to grow the solar system giant planet-cores later due to the low amount of pebbles left. Despite that, we did perform simulations where the bump forms even later, from 0.9 to 1~Myr. This is motivated by the fact that \cite{Kruijeretal17} suggest that Jupiter has to form within 1~Myr to explain the meteorite isotopic dichotomy. The results found for this particular set of simulations are qualitatively similar to those presented in this section, in the scenario where the bump grows from 0.2 to 0.3~Myr.

\begin{figure*}
    \centering
    \includegraphics[scale=0.3]{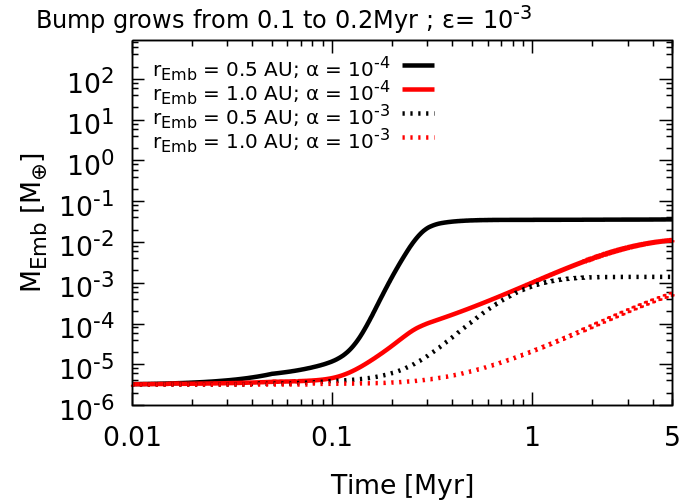}
    \includegraphics[scale=0.3]{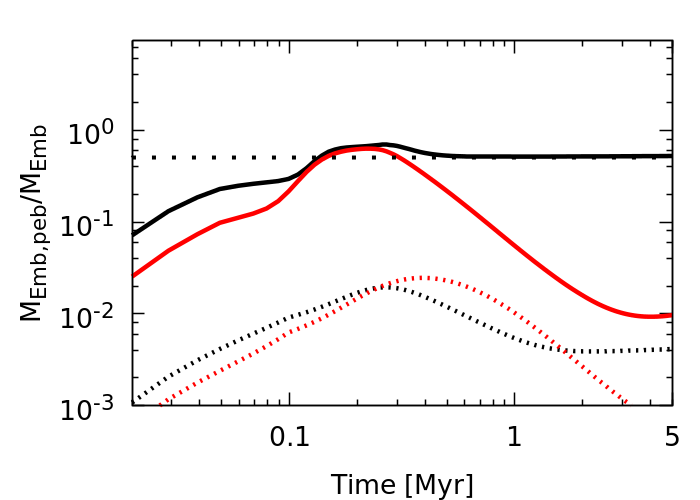}
    \includegraphics[scale=0.3]{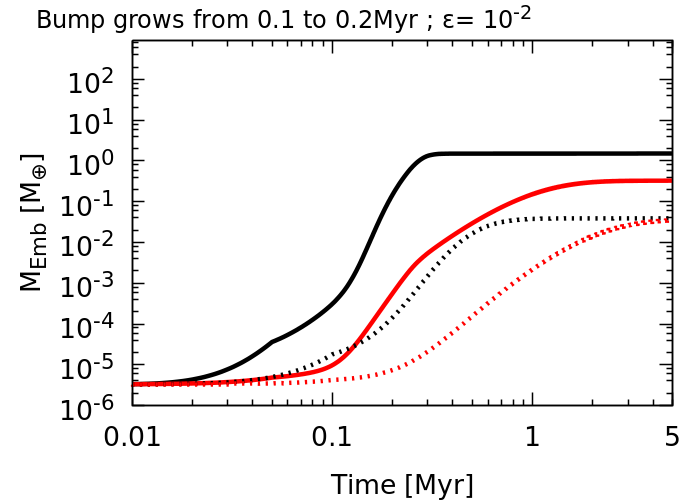}
    \includegraphics[scale=0.3]{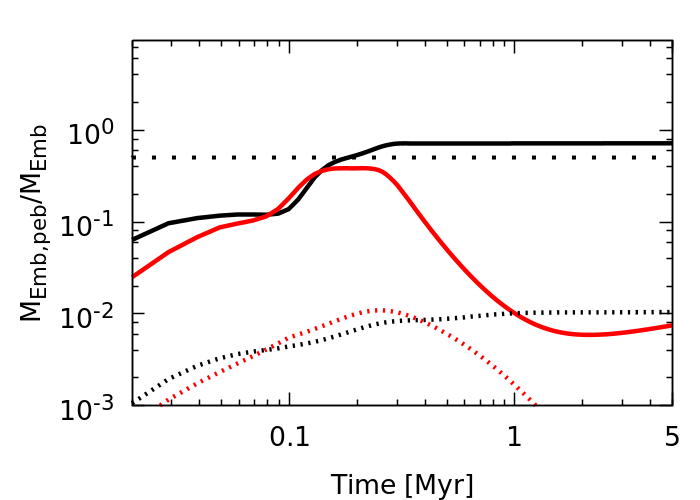}
    \includegraphics[scale=0.3]{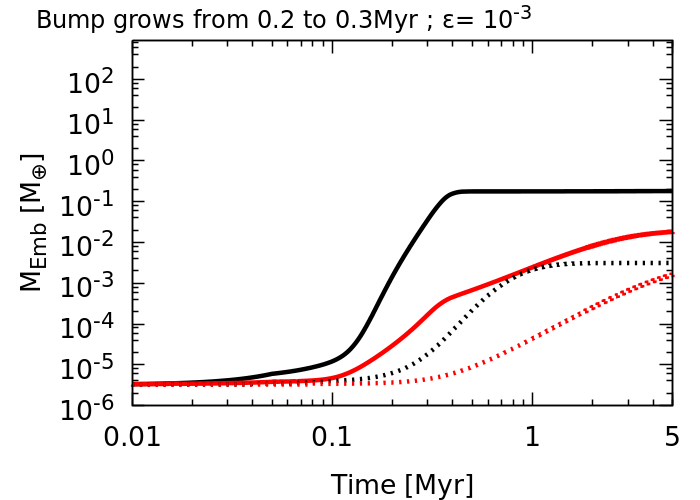}
    \includegraphics[scale=0.3]{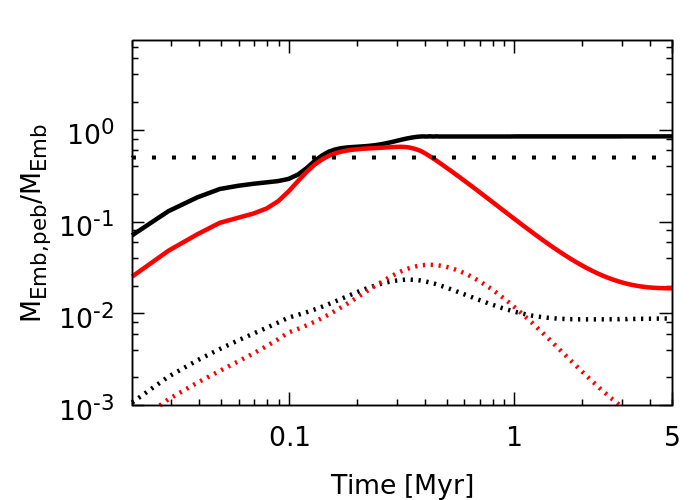}
    \includegraphics[scale=0.3]{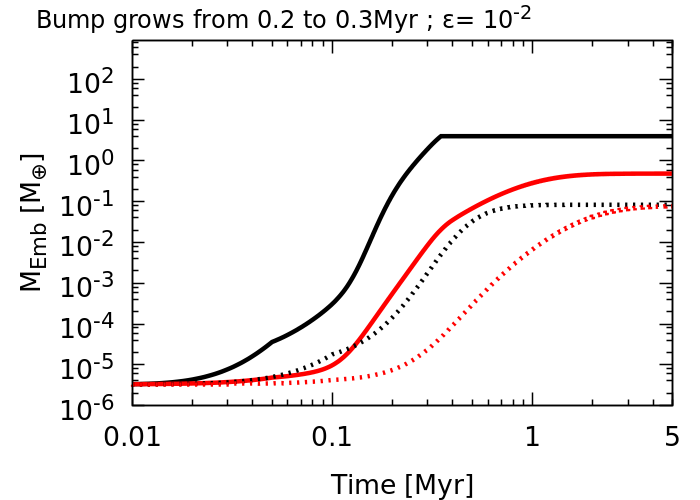}
    \includegraphics[scale=0.3]{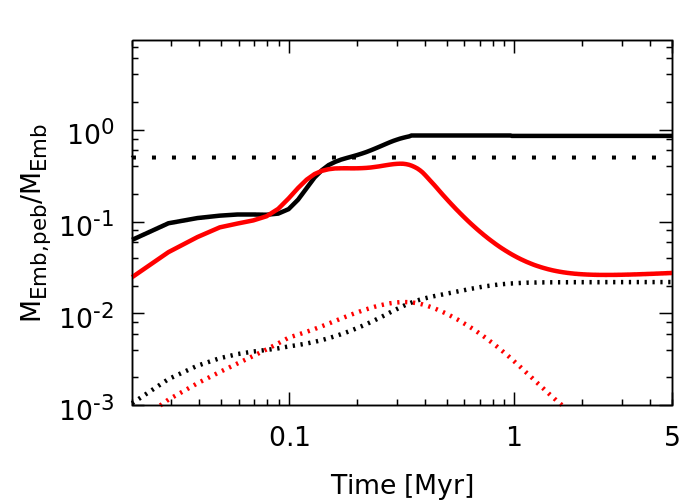}
        \caption{\textbf{Left:}  growth of planetary embryos in simulations where the pressure bump forms at  different times. \textbf{Right:} contribution of pebble accretion to the embryo mass. The horizontal dashed line corresponds to equal contribution from pebble and planetesimal accretion. The time at which the bump starts to form is indicated on the top of each panel together with the bump parameters. In all cases, we use linear interpolation to mimic the bump formation. Solid (dashed) lines show embryos in disks with $\alpha=10^{-4}$, $A=0.5$, and $f=1$ ($\alpha=10^{-3}$; $A=2.0$, and $f=1$). The line colors show the position of the non-migrating embryo.}
            \label{fig:growingbumps}
\end{figure*}

Figure \ref{fig:growingbumps} shows the growth of embryos in simulations where the bump grows at different times. Here, we have performed simulations considering two planetesimal formation efficiencies, namely $\epsilon=10^{-3}$ and $10^{-2}$ and disks with different viscosities. Figure \ref{fig:growingbumps} shows that when $\alpha=10^{-3}$, planetary embryos grow at most to Moon to Mars-mass planetary embryos and that these embryos grow mainly via planetesimal accretion. The model result partially changes for low-viscosity disks. In this case, because pebbles are relatively larger, embryos at 0.5~au quickly grow and start to accrete pebbles before the bump forms and the pebble flux ceases. The contribution of pebble accretion for embryos growing in this setup is typically larger than 50\%. However, we will show later that these embryos form so quickly that they should have migrated inwards, potentially to the disk inner edge ($\sim$0.1~au). Therefore they can not correspond to the terrestrial planetary embryos that formed the solar system. Finally, even in this scenario, where the bump forms later and inner solar system is invaded by a lot pebbles from the outer disk, embryos growing at 1~au, again, grow  relatively more slowly and mostly via planetesimal accretion. Because these embryos grow more slowly they may escape large scale inward migration~($\gtrapprox$1~au; we will address this issue later in Section \ref{sec:mig}).


\subsection{Effects of the threshold fragmentation  velocity inside the snowline}

The threshold fragmentation velocity has a tremendous impact on the size of pebbles and their lifetime in the disk. Here we increase the pebbles sizes that can grow in the inner disk by increasing the threshold fragmentation velocity of pebbles inside the snowline from 1~m/s to 3~m/s. We refer to this quantity as $v_{\rm r,snow}$. Note that this increase in the fragmentation velocity inside the snowline  is  roughly equivalent to assuming that the gas disk midplane has a lower level of turbulence that controls pebble sizes (see Eq. \ref{eq:vfrag}). An increase in the threshold fragmentation velocity by a factor of 3 roughtly correlates to a level of turbulence a factor of $\sim$10 lower in the disk mid-plane (although, in reality, things are a bit more complicated because the pebble disk scale height also depends on the turbulence level).  This view is consistent with gas disk models showing low level of  turbulence in the gas disk midplane~\citep{gammieetal96}. Disk observations also show disks with very low (e.g. $\alpha<10^{-5}$) levels of turbulence~\citep[e.g.][]{dullemondetal18}. Finally, although laboratory experiments suggest that silicate dust grains fragment at velocities of $\sim$1~m/s~\citep{blum18}, dust coagulation models have considered velocities of up 10~m/s~\citep[e.g.][]{birnstieletal10,pinillaetal12}. So we test this scenario for completeness. The fragmentation velocity of ice pebbles is the same of our nominal simulations,i.e., 10~m/s. 

\begin{figure*}
    \centering
    \includegraphics[scale=0.35]{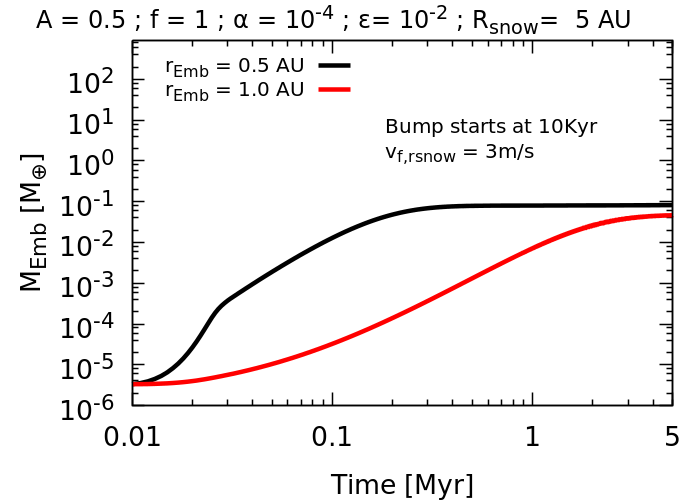}
    \includegraphics[scale=0.35]{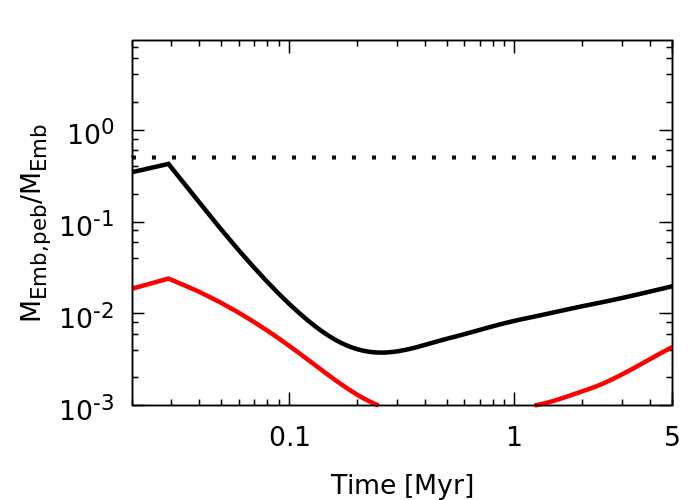}
    
    \includegraphics[scale=0.35]{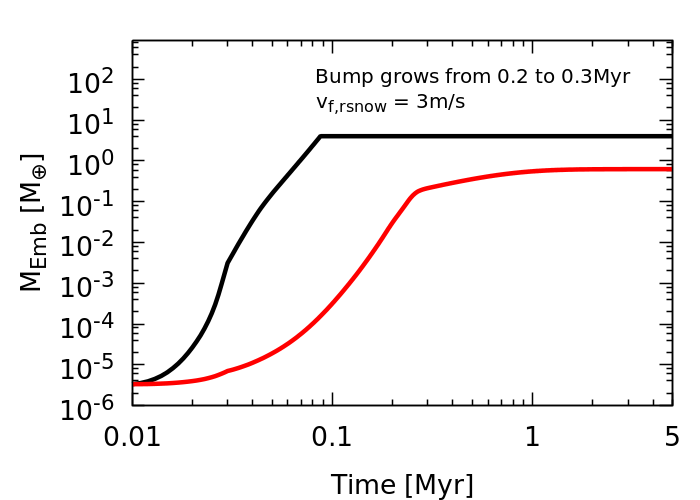}
    \includegraphics[scale=0.35]{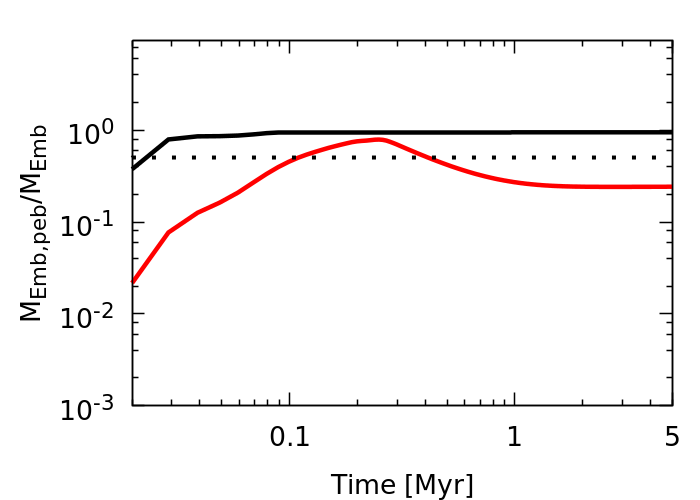}
    \caption{\textbf{Left:} growth of planetary embryos in simulations where the fragmentation velocity inside the snowline is $v_{r,snow}$=3~m/s. \textbf{Right:} Pebble accretion contribution to the embryo's mass. The pressure bump form at different times as indicated on each panel. The disk viscosity, planetesimal formation efficiency, and bump parameters are indicate on the top panel and it is the same in both simulations. As before, in the forming bump simulation, we use linear interpolation to mimic the bump formation. The black (red) line corresponds to an embryo at 0.5~au (1~au).}
        \label{fig:highv_fixgrow}
\end{figure*}

Figure \ref{fig:highv_fixgrow} shows the results of two simulations where $v_{\rm r, snow}$ is higher than our nominal case. In the top-left panel we show the growth of embryos in a disk where the pressure bump is fully formed from the beginning of the simulation and the fragmentation velocity inside the snowline is $v_{\rm r,snow}~=~3$~m/s. In this case, planetary  embryos grow to about Mars-mass and most of their masses comes from planetesimals, similar to our previous results. This is because pebbles in the inner disk are so large (1-10~cm) that they drift inwards very quickly. This effect is indicated on the right-top  panel of  Figure \ref{fig:highv_fixgrow}. First pebble accretion dominates, but at $\sim$0.04~Myr, planetesimal accretion starts to dominate the growth of the embryos because pebbles in the inner region has already drifted inside 0.5~au. In the bottom panels we show the growth of embryos in a disk where the pressure bump starts 0.3~Myr after the beginning of our simulation. Here we also take into account the sublimation of inwawrd drifting ice pebbles crossing the snowline before the bumps forms, assuming a 1:1 ice-to-silicate ratio. In this case, the embryo at 0.5~au grow to a mass larger than the Earth mostly via pebble accretion. The embryo growing at 1~au, grow to a mass larger than 0.5$M_{\oplus}$ in less than half million years. We will show later in the paper that both embryos in this simulation would probably migrate inwards very quickly and become a hot planet/super-Earth~\citep{izidoroetal19,lambrechtsetal19}. Therefore, a system where embryos grow very quickly via pebble accretion in our disk model can not correspond to the current Solar System~\citep[e.g.][]{johansenetal21}.


\subsection{Effects of Type-I migration}\label{sec:mig}

In order to quantify the effects of type-I migration in our previous simulations, we performed several additional simulations to test how
embryos with different masses would migrate in our disk.  Here, we account for the disk evolution by assuming that gas surface density decays exponentially in an e-fold timescale of 1~Myr. We do not model the growth of embryos but instead we assume that they grow to a given mass and we release them to migrate at different times of the disk. Our goal here is not to precisely constraint the migration history of our embryos but only get a sense of their scale of radial migration and how it correlates with their formation time. The prescription of type-I migration considered here is detailed in \cite{izidoroetal17,izidoroetal19} and follow~\cite{paardekooperetal11}. We also impose a planet trap at the disk inner edge, assumed to be at 0.1~au to mimic the effect of the disk inner edge~\citep{romanovaetal03,flocketal19}.
Note that accounting simultaneously for the effects of type-I migration,  pebble, and planetesimal accretion in this work would require N-body numerical simulations, which would be computationally very expensive -- due to the large number of planetesimals required --  and impractical due to the large set of parameters tested in this study. 

\begin{figure}
\centering
    \includegraphics[scale=0.7]{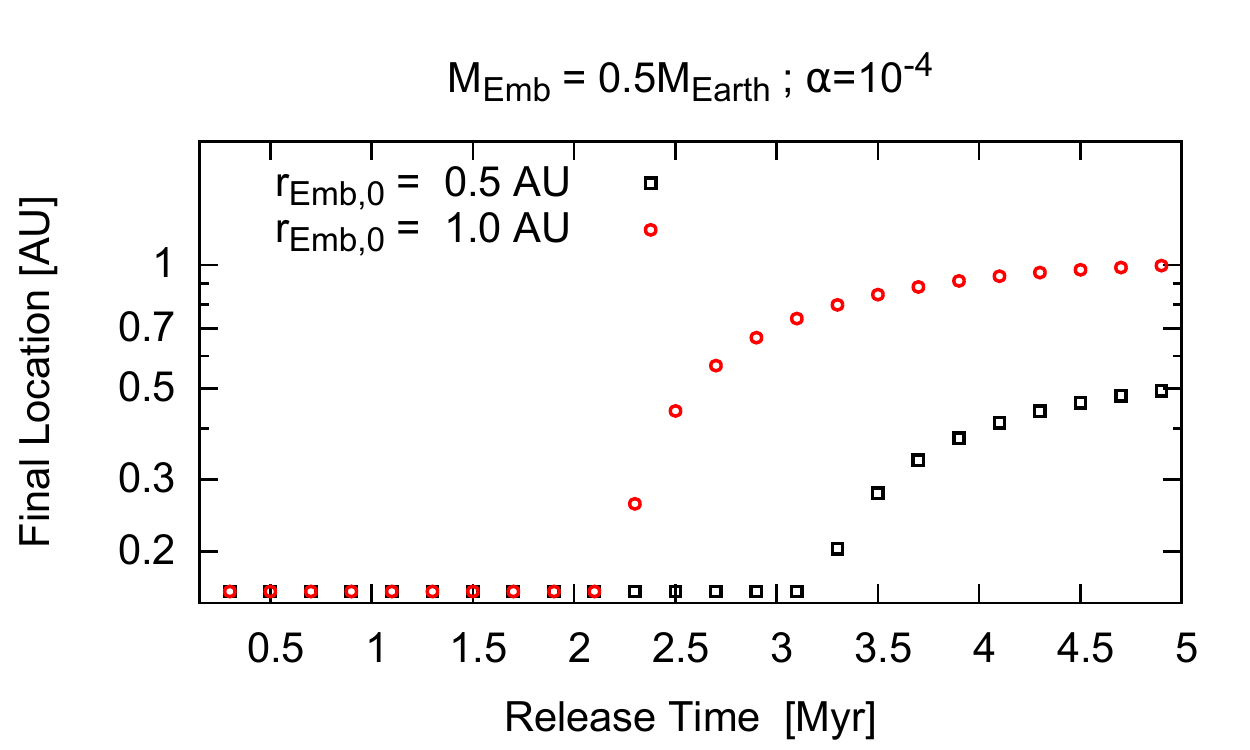}
    \includegraphics[scale=0.7]{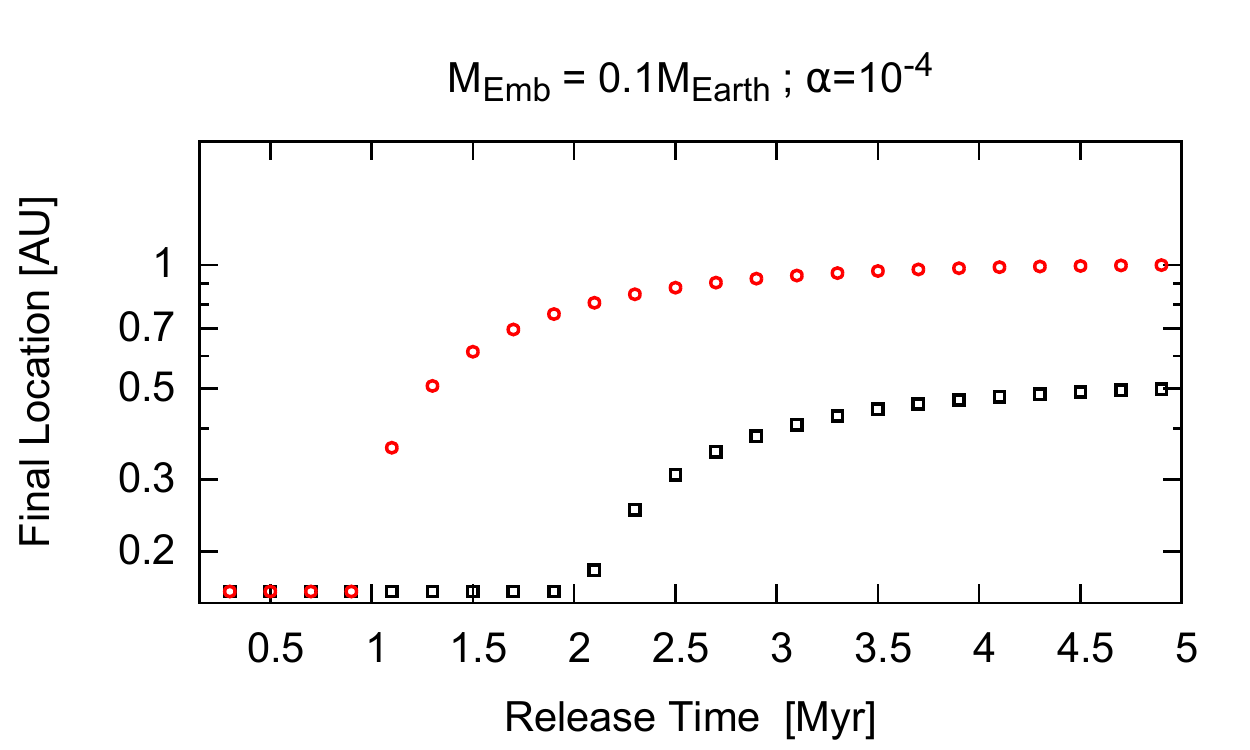}
    \includegraphics[scale=0.7]{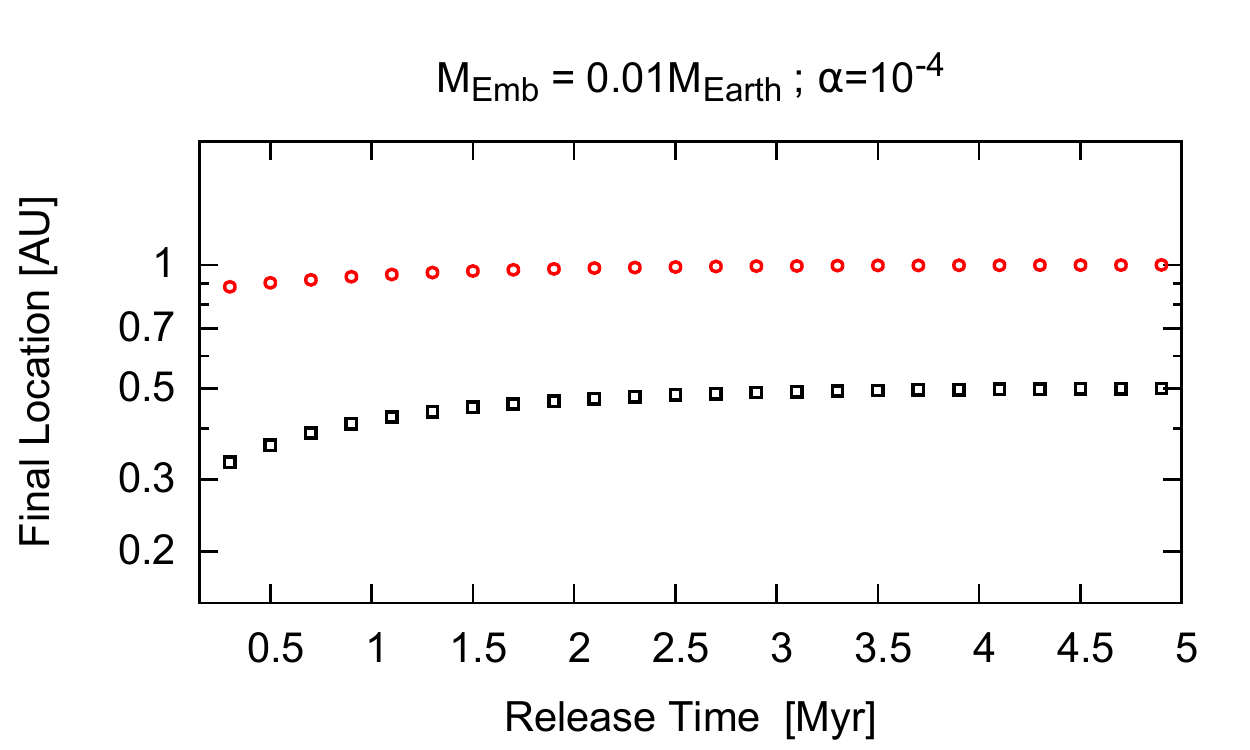}
    \caption{Final location of migrating and non-growing embryos released to migrate at different times of the disk. Each color-coded dot show the final position of an embryo in one simulation. Embryos released from the starting position $r_{\rm Emb,0}=$0.5~au are shown in black, and embryos released from $r_{\rm Emb,0}=$1~au in red. The mass of the embryo and the viscosity of the disk is indicated on the top of each panel.}
    \label{fig:getmigration}
\end{figure}

Figure \ref{fig:getmigration} shows the final location  of embryos released to migrate at different times. Each panel shows, from top to bottom, embryos with masses of 0.5$M_{\oplus}$, 0.1$M_{\oplus}$, and 0.01$M_{\oplus}$. Embryos are released to migrate from 0 to 5~Myr. Figure \ref{fig:getmigration} shows that if an embryo of 0.5$M_{\oplus}$ forms at 0.5~au in less than 3.5~Myr, it should reach the disk inner edge. That would be the case of embryos growing mostly via pebble accretion in Figure \ref{fig:growingbumps} (second panel from top-to-bottom). Even at 1~au, such a massive embryo should not form earlier than  2~Myr. Mars-mass embryos migrate relatively slowly, so our simulations suggest that an embryo with this mass should not reach Mars-mass before 1-2~Myr. Finally, Moon-mass planetary embryos migrate very little and could have formed at any stages of the disk. These results suggest that embryos in the terrestrial region in our disk could not have grown via a vigorous flux of (large) pebbles, otherwise they should have migrated too close to the sun ~\citep{lambrechtsetal19,izidoroetal19}. Figure \ref{fig:getmigration} shows that  avoiding large scale inward migration of our embryos (e.g. those produced in Figure \ref{fig:growthlowvisc} and \ref{fig:growthhighvisc}) requires a sufficiently low planetesimal formation efficiency (e.g. $\epsilon \lesssim10^{-2})$). At the same time, a very small $\epsilon$ may form  low-mass planetary embryos (and potentially terrestrial disk with not enough mass to form all terrestrial planets). Thus, there seems to exist a region of the parameter space where the interplay between growth and migration is just right to make our current solar system. Identifying this  parameter space and how it may change for different disk models remains as subjects for future studies.

One question is whether it is possible to suppress or slow-down type-I migration of these massive terrestrial embryos in our simulations? We discuss two different scenarios that we do not consider in this work. The first one is that -- because our embryos, for simplicity, are assumed to be fully formed in simulations of this section -- we  do not include the effects of thermal torques due to accretion of solids~\cite[]{benitezetal15,masset17}  when computing the embryos' final locations in the disk. Thermal torques may reverse migration of a growing protoplanetary embryo~\cite[see][]{masset17}. We have estimated the critical mass at which the effects of thermal torques are not guaranteed to be efficient enough to promote significant outward migration~\citep[see][]{masset17}. At 1~au, in our model disk, this corresponds to masses larger than about Moon-mass (if the adiabatic index is assume to be 1.4). ~\cite{baumannbitsch20} and \cite{guileraetal19} studied the effects of thermal torques on growing protoplanetary embryos and found similar critical masses.  The effects of disk winds, which trigger mass loss and angular momentum transports in the disk, have also been invoked to suppress inward type-I migration~\citep{ogiharaetal15a,ogiharaetal18,kimmigetal20}. Disk wind effects may induce gas disk surface density profiles with flat or positive slopes in the inner disk~\citep{suzukietal16}. Such profiles may help to halt or even reverse type-I migration in the inner regions of the disk ($\lesssim$1-10~au; \cite{ogiharaetal15}). However, the very gas disk profile depends on assumed disk wind model, with some scenarios not resulting in positive slopes~\citep{bai16}. Inward type-I migration is only fully suppressed in disk-wind simulations if the wind is sufficiently strong (positive gas surface density slope) and  the corotational torque is assumed fully unsaturated. This latter assumption may not be correct for Earth-mass planets and some level of inward migration for planets more massive than Mars may be difficult to avoid~\cite{ogiharaetal18}. Due to the uncertainties of the effects of disk-winds on the migration of low mass planetary embryos, we do not included these effects in our current work.

\subsection{Planetesimal surface density}

In this section, we analyze the planetesimal surface density profiles produced in our simulations with that envisioned in the minimum mass solar nebula model (MMSN; \cite{hayashi81}). Here we  perform an additional set of simulations where we assume that the disk surface density evolves with time (exponential decay with e-fold timescale of 1~Myr), which is certainly more realistic than our initial approach. Also, motivated by the results of 3D hydrodynamical simulations that show that \citep{bitschetal15} the inner parts of protoplanetary disks are not strongly flared, we impose that our disk aspect ratio is constant over radius as  $h(r)=0.037$ in another additional set of simulations. At 10~kyr, pebbles  at 0.5~au in our nominal flared disk with $\alpha=10^{-4}$ are a factor of $\sim$7  smaller than those at 4~au (see Figure \ref{fig:surface_nobump}). In our non-flared disk, this difference is smaller, only a factor of 2.5.

\begin{figure}
    \centering
    \includegraphics[scale=0.35]{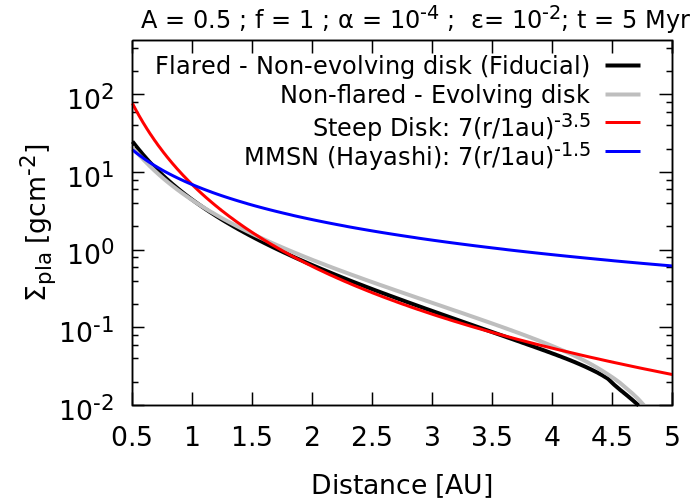}
    \caption{Planetesimal surface density profiles produced in two of our simulations at the end of the gas disk dispersal. A steep surface density profile and the MMSN disk profiles are showed for comparison. For comparison porpuses only we have shown the final planetesimal surface density of a disk where the gas surface density is assumed to dissipate exponentially with an e-fold timescale of 1.0~Myr.}
    \label{fig:planetesimal_profiles}
\end{figure}

The distribution of solids in the MMSN yields a radial surface density profile as $7(r/1{\rm au})^{-1.5}{\rm g/cm^2}$. Figure \ref{fig:planetesimal_profiles} compares the surface density of planetesimals in our simulations with those in the MMSN disk. The disk parameters and pressure bump setup are shown at the top of the Figure. The black solid line shows the final planetesimal surface density in our simulation of Figure \ref{fig:surface_withbump} (left-panels). The gray-line shows planetsimals' surface density in the simulation where the disk is not flared and evolves in time. As one can see, in both disks - flared and non-flared - our final radial distribution of planetesimals are much steeper than that of the MMSN model. Our simulations yields surface density of planetesimals as $\approx~(r/1{\rm au})^{-3.5}$~\citep[see also ][]{lenzetal19}.  Steep surface density profiles may help to alleviate the so called small Mars-problem of classical model of the formation of the terrestrial planets~\citep{izidoroetal15b}. We do not model the late stage of accretion of terrestrial planets in this work, but this result provide an interesting avenue for future research.

\section{Discussion} \label{sec:discussion}

Our model is build on a series of assumptions and  simplified in many aspects. For instance, we do not  self-consistently model the viscous evolution of the disk~\citep[e.g.][]{birnstieletal10} and the formation of the pressure bump taking into account thermodynamical effects in the disk. Our simulations also  follow the evolution of only two dust populations rather than a continuous of dust sizes~\citep{birnstieletal10}. Our pressure bump parameters are deliberately chosen to provide an efficient separation of the inner and outer dust/pebble reservoirs in a 1D-model. Bi-dimensional hydrodynamical simulations, however, show that sufficiently small dust particles can still cross pressure bumps created by gap-opening planets~\cite[e.g.][]{weberetal18,tamfaletal18,bitschetal18b,
haugbolleetal19,survilleetal20}. It is unclear how much material in the form of small dust grains -- potentially with CC-like composition -- entered into the inner Solar System during the sun's natal disk phase. It is unclear how much of this dust would be able to participate in the formation of planetesimals, for instance, via streaming instability~\cite[see also][]{schilleretal18} and/or significantly contribute to the formation of planetary embryos via pebble accretion.

When calculating the velocity of the gas we have neglected the back-reaction of the dust on the gas~\citep[e.g.][]{drazkowskaetal16}. We have performed some simulations to quantity these effects and they do not affect qualitatively our main conclusions. When modeling planetesimal formation and embryo growth, we have also neglected the self-regulation effect that planetesimals and embryos may have on the formation of planetesimals~\cite[e.g.][]{drazkowskadullemond14}. 

\begin{figure}
    \includegraphics[scale=0.35]{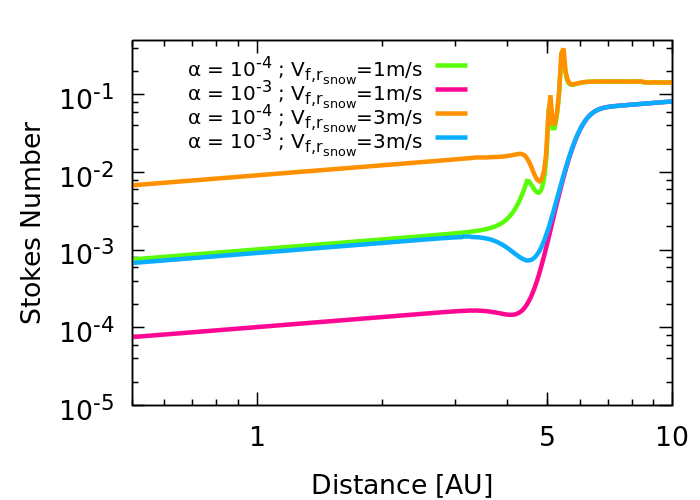} 
        \caption{Stokes number of pebbles at the beginning of our simulations with different disk setups and fragmentation velocities. Our simulations start at t=10~Kyr so pebbles have already grown inside $\sim$10~au. Sharp variations observed in these curves are due to the pressure bump location and snowline.}
            \label{fig:stokes}
\end{figure}

Another important assumption in our work is that we assume that pebbles with Stokes number larger than $5\times10^{-5}$ can concentrate and collapse to form planetesimals~\cite[see also][]{voelkeletal20}. Figure \ref{fig:stokes} shows the stokes number of the pebbles in the inner disk at the beginning of our simulations. If zonal flows, vortices, and  the effects of streaming instability can only efficiently concentrate pebbles at levels required for planetesimal formation, i.e., if St$>10^{-3}-10^{-2}$~\citep{raettigetal15,yangetal17,simonetal16,carreraetal17,drazkowskaalibert17,klahretal18,lenzetal19}, then planetesimals formation would not happen inside the disk snowline in our disks with $\alpha=10^{-3}$ and $v_{\rm f,rnow}=$1~m/s (pebbles fragmentation velocity inside the snowline). Disks with $\alpha=10^{-4}$ and $v_{\rm f,rsnow}$=1~m/s  on the other hand, would form planetesimals only beyond $\sim$0.8-1~au. That could be an alternative way to explain why the Solar System did not form planetesimals inside 0.5~au and explain the current location of Mercury and the lack of close-in super-Earths in our planetary system~\citep[see also][]{izidoroetal15a,izidoroetal15c,johansenetal21}. 

Our simulations start from a fully formed disk. Recent simulations have suggested that processes occurring during the early stages of formation of the gas disk itself are important to explain the isotopic dichotomy between NC and CC meteorites~\cite[]{lichtenbergetal21}.
Our model differs from their in many ways. For instance, the simulations of \cite{lichtenbergetal21} invoke very different levels of turbulence controlling the growth of pebbles in the disk midplane (they refer to it as $\alpha_t=10^{-5}$) and the transport of angular moment in the gas disk (they refer to it as $\alpha_{\nu}=10^{-3}$) to allow prompt formation of planetesimal in the inner disk during the disk expansion~\cite[see also][]{pinillaetal21}. In most of our simulations, $\alpha_t=\alpha_{\nu}$, and in our most extreme scenarios $\alpha_t$ would corresponds a value at most 10 times smaller than $\alpha_{\nu}$. The planetesimal formation model of \cite{lichtenbergetal21} is also different from ours. Planetesimal formation only takes place in their simulations if the ratio between the pebble column density and gas column density in the disk midplane is larger than $\sim$1. Here we assume that zonal flows and vortices help concentrate pebbles and trigger planetesimal formation virtually everywhere in the disk~\citep[e.g.][]{lenzetal19}.

We have also performed simulations where the disk is hotter and the snowline is further out, at 10~au, rather than 5~au as in our nominal simulations. We do not present these results here because even in this extreme scenario they are qualitatively similar to those of our nominal simulations.

Although our model is admittedly simple, it also provides valuable insights on planet formation. We show that without a pressure bump, the inner solar system would be quickly invaded by several tens to a few hundreds of Earth masses in pebbles from the outer disk~\cite[see also][]{lichtenbergetal21}. A large pebble flux may complicate the evolution of the inner solar system not only in terms of not accounting for the isotopic dichotomy between the NC and CC reservoirs but also affecting the final masses and orbits of the planets. Our model also suggest that if the inner and outer solar system pebble reservoirs were early and efficiently disconnected  terrestrial planetary embryos formed via planetesimal accretion, and terrestrial planets completed accretion most likely via collisions of Moon to Mars-mass planetary embryos. Interior modeling of planetary evolution may help to distinguish different modes of planetary accretion, namely either pebble or planetesimal dominated growth. It is interesting to note that some of our simulations form Earth-mass planets mostly via pebble accretion at 0.5~au but never at 1~au. Embryos at 1~au typically grow  to lower masses mostly via planetesimal accretion which require mutual collisions of embryos to produce an Earth-mass planet at 1~au. A scenario where a Venus-mass embryo forms at $\sim$0.5~au via pebble accretion and somehow avoids inward gas-driven migration and a phase of giant impacts could be an interesting avenue to explain why Venus looks so different from Earth.

In this work we do not model the fate of  pebbles that reach the disk inner edge and the movement of the pressure bump as the disk evolves. We also do not model the growth of cores in the bump, potentially mirroring Jupiter's core growth~\cite[e.g.][]{morbidelli20,guilleraetal20}. This remains as topics of  future studies.

\section{Summary} \label{sec:conclusions}

In this paper we use a 1D dust evolution code to model the impact of a pressure bump on the formation of terrestrial planetary embryos in the solar system. Our model includes the effects of radial drift, coagulation, fragmentation, and turbulent mixing of dust grains. We also model planetesimal formation and planetary embryo's growth via planetesimal and pebble accretion. In our simulations, we assume that a pressure bump formed early in the disk -- potentially accelerating the growth of Jupiter -- lead to an efficient separation of  the inner (NC-like) and outer (CC-like) solar system pebble reservoirs. In this scenario, our simulations show that terrestrial planetary embryos most likely grew via planetesimal accretion rather than pebble accretion. When the pressure bump disconnects the inner and outer solar system, pebbles in the inner disk drift inwards very quickly due to gas drag. Pebble drift foster planetesimal formation  via collapse of over-dense pebble clumps -- if pebbles can get efficiently concentrated -- but  most pebbles are lost at the disk inner edge before planetary embryos can grow to large enough masses where pebble accretion becomes very efficient. Consequently, planetary embryo's growth around $>$1~au is dominated by planetesimal acretion. Only if $\gtrapprox100M_{\rm Earth}$ in  pebbles from the outer disk drift into the inner solar system before the formation of the bump (at $\sim$0.2-1~Myr) and planetesimal formation efficiency is sufficiently low, pebble accretion dominates the growth of embryos at about $<$0.5-1~au. However, these embryos become typically so massive and/or form so quickly that they should move inwards, by type-I migration, interior to Mercury's orbit, which is inconsistent with the current solar system architecture. We have demonstrated that our results remain qualitatively valid for a  plausible large part of the parameters space. In order for pebble accretion to dominate the formation of terrestrial planetary embryos, fine-tunning of initial conditions would be required. Finally, all of our simulations always form planetesimals inside 0.4-0.5~au in the solar system. We envision two potential solutions for this issue that should be further investigated: i) either our criteria for planetesimal formation to occur is too generous and, in reality, planetesimal formation never took place in the innermost parts of the disk due to low stokes number and/or low dust-to-gas ratio; or ii) the solar system disk was much hotter than our nominal disk model and the silicate sublimation line was at about 1~au when the outer and inner solar system was separated.

\bibliography{sample63}{}
\bibliographystyle{aasjournal}

\acknowledgments
We thank the anonymous reviewer for providing valuable comments and suggestions that helped improving the quality of our original manuscript. A.~I. is truly grateful to Rogerio Deienno, Sean Raymond, Andrea Isella, Damanveer Grewal, and Akash Gupta for inspiring discussions and motivation during the development of this work. A.~I and R.~D.  acknowledge NASA grant 80NSSC18K0828 for financial support during preparation and submission of the work. B.B., thanks the European Research Council (ERC Starting Grant 757448-PAMDORA) for their financial support.  

\end{document}